\documentclass[12pt,preprint]{aastex}

\newcommand{\pks}{PKS~0637$-$752}
\newcommand{\lae}{\mathrel{\raise .4ex\hbox{\rlap{$<$}\lower 1.2ex\hbox{$\sim$}}}}
\newcommand{\gae}{\mathrel{\raise .4ex\hbox{\rlap{$>$}\lower 1.2ex\hbox{$\sim$}}}}

\slugcomment{Submitted to Ap. J. Supplements}
\shorttitle{X-rays from Quasar Jets}
\shortauthors{Marshall et al.}

\begin{document}

\title{An X-ray Imaging Survey of Quasar Jets -- Testing the Inverse Compton Model}

\author{H. L. Marshall\altaffilmark{1},
J.M. Gelbord\altaffilmark{1,8,9},
D.A. Schwartz\altaffilmark{2},
D.W. Murphy\altaffilmark{3},
J.E.J. Lovell\altaffilmark{4,10},
D.M. Worrall\altaffilmark{2,5},
M. Birkinshaw\altaffilmark{2,5},
E. S. Perlman\altaffilmark{6,11},
L. Godfrey\altaffilmark{3,7},
D.L. Jauncey\altaffilmark{3}}
\altaffiltext{1}{Kavli Institute for Astrophysics and Space Research,
 Massachusetts Institute of Technology, 77 Massachusetts Ave.,
 Cambridge, MA 02139, USA}
\altaffiltext{2}{Harvard-Smithsonian Center for Astrophysics,
 60 Garden St., Cambridge, MA 02138, USA}
\altaffiltext{3}{Jet Propulsion Laboratory, 4800 Oak Grove Dr.,
Pasadena, CA 91109, USA}
\altaffiltext{4}{CSIRO Australia Telescope National Facility,
PO Box 76, Epping, NSW 2121, Australia}
\altaffiltext{5}{Dept. of Physics, University of Bristol, Tyndall Ave.,
Bristol BS8 1TL, UK}
\altaffiltext{6}{Joint Center for Astrophysics and Physics Department, Univ. of Maryland,
 Baltimore County, 1000 Hilltop Circle, Baltimore, MD, 21250, USA}
\altaffiltext{7}{Curtin Institute of Radio Astronomy, Curtin University of
 Technology, Bentley, WA 6845, Australia}
\altaffiltext{8}{Dept. of Physics, Durham University, South Rd., Durham, DH1 3LE,UK}
\altaffiltext{9}{Dept. of Astronomy and Astrophysics, Pennsylvania State University,
State College, PA, 16801, USA}
\altaffiltext{10}{School of Mathematics and Physics, University of
 Tasmania, Hobart, TAS 7001, Australia}
\altaffiltext{11}{Dept. of Physics and Space Sciences,
 Florida Institute of Technology, 150 W. University Blvd., Melbourne, FL, 32901, USA}
\email{hermanm@space.mit.edu,
jgelbord@astro.psu.edu,
das@head-cfa.harvard.edu,
david.murphy@jpl.nasa.gov,
jim.lovell@utas.edu.au,
D.Worrall@bristol.ac.uk,
Mark.Birkinshaw@bristol.ac.uk,
eperlman@fit.edu,
L.Godfrey@curtin.edu.au,
David.Jauncey@csiro.au
}

\slugcomment{Accepted for publication in the Ap J Supplements}

\begin{abstract}

We present results from continued
{\em Chandra} X-ray imaging
and spectroscopy of a flux-limited sample
of flat spectrum radio-emitting quasars with jet-like
extended structure. 
X-rays are detected from 24 of the 39 jets observed so far. 
We compute the
distribution of $\alpha_{rx}$, the spectral index between the
X-ray and radio bands, showing that it is broad, extending at
least from $0.8$ to $1.2$.
While there is a general
trend that the radio brightest jets are detected most often, it is clear that
predicting the X-ray flux from the radio knot flux densities is risky so a shallow X-ray
survey is the most effective means for finding jets that are X-ray bright.
We test the model in which the X-rays result from inverse Compton (IC)
scattering of cosmic microwave background (CMB) photons by
relativistic electrons in the jet moving with high bulk Lorentz
factor nearly along the line of sight.
Depending on how the jet magnetic fields vary with $z$,
the observed X-ray to radio flux ratios do not follow the
redshift dependence expected from the IC-CMB model.
For a subset of our sample with known superluminal motion
based on VLBI observations, we estimate the angle of
the kpc-scale jet to the line of sight by considering the
additional information in the
bends observed between pc- and kpc-scale jets.
These angles
are sometimes much smaller than estimates based
on the IC-CMB model with a Lorentz factor of 15,
indicating that these jets may decelerate
significantly from pc scales to kpc scales.
\end{abstract}


\section{Introduction}

Many fundamental physical properties of quasar jets remain uncertain,
such as the nature of the energy-carrying particles, whether the
particle energy densities are in equipartition with the local magnetic
field energy densities, and how much entrainment there is.
From the
observation of superluminal motion with the VLBI technique, it is
generally agreed that the pc-scale jets of high power quasars are
highly relativistic, with bulk Lorentz factors ($\Gamma$) of
10-30.
However, it is not certain whether most jets at the kpc-scale
also have high Lorentz factors in bulk motion and whether the jets
are oriented close to our line of sight, as inferred
for \pks\ \citep{celotti,tavecchio} because of its X-ray
bright knots \citep{schwartz}.  The model posited by Celotti et al.\
and Tavecchio et al.\ involved inverse Compton scattering
of photons from the Cosmic Microwave Background (IC-CMB),
in contrast to earlier synchrotron and synchrotron self-Compton
models.
For a review of relativistic jet physics and the role of X-ray
observations, see
\citet{2009A&ARv..17....1W} and references therein.

It is now becoming evident that the simplest, single-zone IC-CMB model
is inadequate in many cases \cite[e.g.][]{ks05,hardcastle06,2006ApJ...648..900J,aneta2}.
One concern with this model is that the lifetimes of the
electrons responsible for the X-ray emission are orders of
magnitude longer than those producing the radio emission so the observed
correspondence of radio and X-ray structures would not be expected
\citep{tavecchio03,schwartz06}.
Extra synchrotron components are proposed by others \citep{2006ApJ...648..900J,hardcastle06}.
In some cases it gets difficult to generate an adequate
physical model \citep{aneta2}.
Currently, the field is in a fruitful phase 
of mutually driven theoretical and observational advances.
Solutions seem to be as varied as the sources themselves,
bolstering the need for more detailed case studies.
This need provides the primary motivation for our
X-ray imaging survey with {\em Chandra} \citep[][hereafter, Paper I]{marshall05}.
Our survey is similar to that undertaken by
\citet{sambruna02,sambruna04} but
the sample is somewhat larger and the
exposures correspondingly shorter.  This being a shallow survey,
we leave detailed modeling of individual sources to later,
follow-up analyses of deeper observations

This paper is a continuation of Paper I and presents
observations of another 19 quasars from the original sample
of 56.  We describe the sample
properties in section 2.  In section 3, we
describe the {\em Chandra} observations and compare
the X-ray maps to newly obtained radio images.  In section
4, we examine the sample properties in the context of
beaming emission models and test the IC-CMB model
in a limited context, as examined previously
by \citet{2004ApJ...600L..23C} and \citet{ks05}.
We use a cosmology in which $H_0 = 70$ km s$^{-1}$ Mpc$^{-1}$,
$\Omega_{\rm m} = 0.3$, and $\Omega_\Lambda = 0.7$.

\section{Sample Properties}

Sample selection was described in Paper I.  Briefly,
56 sources were selected from 1.5 or 5 GHz VLA and ATCA imaging
surveys \citep{murphy,lovell}.  The dominant selection criterion is on
core flux density -- as applied when creating the samples for
the radio imaging surveys.
We then use the flux density in extended
emission, determined from imaging studies of
the sample, as the primary criterion for inclusion in our sample.  
A few sources
have somewhat indistinct morphology but most have
double or triple structure and many have linear
structure.
Subsamples were defined in Paper I:
the ``A'' list was a purely jet flux-limited sample, while the
``B'' list was selected on the basis of morphology
with a bias toward one-sided and linear structure.
The sample had 28 objects in each list, although many
objects from the ``A'' list could qualify morphologically
for the ``B'' list.

So far, 39 sources in the sample have been observed
with {\em Chandra}.  We reported results for the first 20
targets in Paper I, finding that 60\% of the jets could
be detected in short {\em Chandra} exposures.
Here, we present results for another 19
quasars in the sample and present some ensemble
properties for the 39 that have been observed
with {\em Chandra}.  Ten of the new images were obtained
as part of the continuation of our survey and the other
nine were taken from the {\em Chandra} archive.  Some of these
archival observations have somewhat longer exposures than we
have used in our survey.

A significant fraction of the sample has been observed with
VLBI: 22 of the 29 northern sources are in the MOJAVE
survey\footnote{See the MOJAVE web
page: {\tt http://www.physics.purdue.edu/astro/MOJAVE/} and
\citet{2009AJ....138.1874L}.}.
An additional 4 targets with declinations in the
$-40$\arcdeg\ to $-20$\arcdeg\ range have also been observed.
Superluminal motions have been detected and measured for 22 of these
26 quasars; the distribution of the apparent velocities, $c \beta_{app}$,
is comparable to those of the remaining MOJAVE sources,
indicating that our sample and the MOJAVE sample have
similar characteristics.

\section{Observations and Data Reduction}

\subsection{Imaging results}

The observation dates and
exposure times for the {\em Chandra} observations used in this
analysis are given in Table~\ref{tab:observations}.
As in Paper I,
events in the 0.5-7.0 keV range were selected for all analysis and
to form the X-ray images, shown in Fig.~\ref{fig:images}.
The images of the brightest sources show readout streaks, which
were avoided by selecting a suitable range of observatory roll
angles.

Radio maps were obtained for all of the sources at the ATCA
or the VLA archives.
ATCA observations will be reported in detail in a separate paper.
These maps were used for the image overlays in Figure~\ref{fig:images}
and Table~\ref{tab:radioCont} gives a log of the radio observations.
We used these maps to determine radio flux densities for the jets.
Images were registered as in Paper I.

We tested for the detection of X-rays from a jet using a simple Poisson test, as
in Paper I, for counts in a rectangular region extending over a specific
angular range ($\theta_i$, $\theta_o$) from the core at
a specific position angle, and appropriate width.
The radio images were used to define the position angles and lengths
of possible jets.  Most jets are clearly defined as one-sided structure
but in a few ambiguous cases, the pc-scale images were used to define
the jet direction, when available.
The parameters of the selection regions are given in Table~\ref{tab:jetresults}.
The width of the rectangle was
3\arcsec\ except for 0234$+$285, 0454$-$463,
1055$+$018, 1055$+$201, 1928$+$738, and 2007$+$777, where the
jets bend substantially, so the
rectangles were widened up to 4-10\arcsec.
Profiles of the radio emission along the jets are shown in
Fig.~\ref{fig:radioprofiles}.  In order to eliminate X-ray counts
from the wings of the quasar
core, a profile was computed at 90\arcdeg\ to the jet and subtracted.
The X-ray counts in the same rectangular region defined by the radio data
were compared to a similar sized region on the opposite side of
the core for the Poisson test.
We set the critical probability for detection of an X-ray jet
to 0.0025, which yields a 5\% chance that there might be one
false detection in a set of 20 sources.
Histograms of the X-ray emission along the jets are shown in
Fig.~\ref{fig:xrayprofiles}.  The jet and counter-jet position angles
are compared, providing a qualitative view of the X-ray emission along
the jets.  In no case is a counter-jet apparent in the X-ray images.

Jet X-ray flux densities (Table~\ref{tab:jetresults})
were computed from count rates using the conversion factor
1 count/s $=$ 1 $\mu$Jy.  This
conversion is accurate to within 10\% for typical power law spectra.
The spectral index from radio to X-ray is computed using $\alpha_{rx} = 
-\log(S_x/S_r) / \log(\nu_x/\nu_r)$, where $\nu_x = 2.42 \times 10^{17}$ Hz
and $\nu_r$ depends on the map used.
Due to the wide range of redshifts for the observed sample of 39 sources,
the apparent 0.5-7.0 keV X-ray luminosities of the detected
jets range from $10^{40}$ erg/s to
over $8 \times 10^{44}$ erg/s, with a
median value of $8 \times 10^{43}$ erg/s.
Excluding three sources with $z < 0.1$, the minimum detected
jet luminosity is $9 \times 10^{42}$ erg/s and the median
is $1.3 \times 10^{44}$ erg/s.

Redshifts are still unknown for two objects in the sample
for which we have X-ray images: PKS 1145$-$676
and PKS 1251$-$713.  We excluded these two sources
from sample analyses that require redshifts.

\subsection{Notes on Individual Sources}
\label{sec:sources}

In this section, we present qualitative descriptions of the X-ray and
radio morphologies shown in Figure~\ref{fig:images} and describe the
directions of any pc scale jets.  Profiles of the 
radio and X-ray emission along the jets are given in Figures~\ref{fig:radioprofiles}
and \ref{fig:xrayprofiles}, respectively.   All position angles (PAs) are
defined as positive when east of north with due north defining zero.

\subsubsection*{0234$+$285 (4C +28.07)}

On the pc scale,VLBI observations show two jet knots with PAs of about
-15\arcdeg\ and apparent speeds of about 12$c$ \citep{2009AJ....138.1874L}.
The VLA image shows a jet about 6\arcsec\ long with an initial direction of due north but
curving to a PA of
-20\arcdeg\ after which it bends sharply to a PA of -90\arcdeg,
and terminates within 3\arcsec\ at a bright component.
X-ray emission is clearly detected up to the sharp bend.
There is a marginal X-ray detection near the radio feature at the end of the detected jet.

\subsubsection*{0454$-$463 (PKS B0454-463)}

The ATCA data show hotspots placed somewhat asymmetrically 5.9\arcsec\ to
the southeast and 4.5\arcsec\ to
the northwest of the core.  A 3\arcsec\ long
jet extends to the southeast before bending to the south hotspot.  There is
no apparent VLBI structure \citep{2004AJ....127.3609O}.  The X-ray source
appears extended along the jet direction to the southeast, coincident with the
jet before it bends south.
An X-ray source is found 1\arcsec\ beyond the southeast hotspot.

\subsubsection*{0820$+$225 (4C +22.21)}

Extended radio emission surrounds the source, but is oriented predominantly
along an east-west axis.  A VLBI observation by \cite{2000MNRAS.319.1125G}
shows an S-shaped jet extending over 20 mas generally to the southwest.
We consider the jet to be oriented due west
for the purposes of this analysis because the radio emission
is somewhat brighter in this direction on a scale of a few arcsec.
We detect no associated extended X-ray emission.

\subsubsection*{0923$+$392 (4C +39.25)}

VLBA data show knots to the west of the core at a PA of -78\arcdeg,
moving with a maximum apparent speed of 2.9$c$ \citep{vlba,ls00}.
\cite{2009AJ....138.1874L}, however, place the core differently (at position S instead
of A in the \cite{ls00} map, as suggested by \cite{1997A&A...327..513A}), with knots
at a PA of about 100\arcdeg, moving at up to 4.3$c$.
The 4.95 GHz map shows a 2\arcsec\ long
jet extending along a PA of $+$75\arcdeg,
which we take as the direction to examine for X-ray emission.
We detect no associated extended X-ray emission, on either side
of the core.

\subsubsection*{0954$+$556 (4C +55.17)}

This X-ray image was first published by \cite{2007ApJ...662..900T}
and is included in our analysis for completeness.
The VLA image shows two distinct features, one at 5\arcsec\ from the 
core along a PA of -60\arcdeg, and the other at
a PA of 45\arcdeg, about 2.5\arcsec\ from the core.
A MERLIN map at 5 GHz shows a jet at a PA of about
-60\arcdeg\ extending 150 mas to the WNW \citep{1995ApJS...99..297X}
and a 5 GHz map with the VLA shows a 5\arcsec\ long jet at the same
PA as well as the knot 2.5\arcsec\ from the core at a position
angle of 45\arcdeg.
The 5\arcsec\ feature is detected in the X-ray image and is the more likely to
be associated with a jet.

\subsubsection*{1040$+$123 (3C 245)}

This source was in the \cite{sambruna04} sample
and is included in our analysis for completeness.
We detect no significant extended X-ray emission associated with the
western radio emission, which is the direction of
the kpc scale jet as well as optical knots \citep{sambruna04}.
The eastern lobe appears to be detected, so
our algorithm for detecting an X-ray excess on the jet side
(by comparing the X-ray profile along the position
angle of the jet to a region 180\arcdeg\ from it) may be failing
in this case because the existence of what appears to be
inverse Compton emission from the brighter radio lobe \citep[see also][]{sambruna04}.

\subsubsection*{1055$+$018 (4C +01.28)}

This source was in the \cite{sambruna04} sample.
It is included in our analysis for completeness.
This source is another case where the pc-scale jet is
strongly misaligned with the kpc-scale jet.
Jet components are found from 1-10 mas from
the core along a PA of -50\arcdeg, moving at up to 11$c$
\citep{2009AJ....138.1874L,2009ApJ...706.1253H}.
The kpc scale jet is oriented generally to the south
\citep[][Fig. 1]{murphy}.
We detect no extended X-ray emission associated with
the radio emission on a few arcsec scale.
The readout streak is very strong in the image shown
in Fig.~\ref{fig:images}.

\subsubsection*{1055$+$201 (4C +20.24)}

The radio and X-ray images show emission on a 20\arcsec\
scale to the north of the core.  A detailed discussion of this source
has been presented by \cite{schwartz06}.

\subsubsection*{1116$-$462 (PKS B1116$-$462)}

The ATCA image shows a knot about 3\arcsec\ to the west of the core.
We detect no associated extended X-ray emission.

\subsubsection*{1251$-$713 (PKS B1251$-$713)}

The ATCA image shows a knot about 10\arcsec\ to the south of the core.
VLBI at 8.64 GHz shows no structure on the mas scale \citep{ojha05}.
We detect no associated extended X-ray emission.

\subsubsection*{1354$+$195 (4C +19.44)}

This source was in the \cite{sambruna04} sample.
The radio/X-ray jet is over 20\arcsec\ long and was the subject
of a follow-up observation, which is used in a more in-depth
analysis \citep[see][and Harris et al., in prep.]{2007Ap&SS.311..341S}.
It is included here for completeness.
The readout streak is very strong in the image shown
in Fig.~\ref{fig:images}.
Based on VLBA observations from the 2 cm survey \citep{vlba},
the pc-scale jet of 1354$+$195 is relatively straight, at a position angle of 143\arcdeg.
Several bright jet features appear to
have motions of 160-240 $\mu$arcsec/yr (6.6-9.9 $c$)
(Lister et al.\  in prep.).

\subsubsection*{1421$-$490 (PKS B1421$-$490)}

This observation was published by
\cite{gelbord05} and is included here
for completeness.  Magellan spectra show that component B
(in the center of the image shown in Fig.~\ref{fig:images}), which 
is the brightest optically and in the X-ray band, is the core of the quasar with $z = 0.662$.
Because component A (to the northeast of component B and brighter in the radio band)
has been resolved using VLBI
\citep{godfrey08}, we now consider that component
B is associated with the quasar core and that component A is a
radio lobe with an exceptionally bright and compact hotspot.

\subsubsection*{1641$+$399 (3C 345)}

This source was in the \cite{sambruna04} sample
and is included in our analysis for completeness.
VLBA data show pc scale jet knots moving at
about 12$c$ along an average
PA of -90\arcdeg\ \citep{vlba,2009AJ....138.1874L}.  The jet
curves north to a PA of about -45\arcdeg\
about 4 mas from the nucleus \citep{2009AJ....138.1874L}.
The kpc scale images (Fig.~\ref{fig:images}) show
an X-ray and radio knot about 3\arcsec\ from the core
at a PA of -35\arcdeg.

\subsubsection*{1642$+$690 (4C +69.21)}

This source was in the \cite{sambruna04} sample
and is included in our analysis for completeness.
The pc-scale jet is oriented along a position
angle of -162\arcdeg\ and has a maximum apparent
speed of 16$c$ \citep{vlba} while the kpc-scale jet
points due south and curves to the east \citep{odea88}.
There appears to be an X-ray excess associated with the
brightest part of the extended jet, $\sim 3$\arcsec\ from the core.

\subsubsection*{1928$+$738 (4C +73.18)}

This source was in the \cite{sambruna04} sample
and is included in our analysis for completeness.
The VLBA data show many knots moving non-radially, at 2-8$c$
on an arc that extends over PAs 150-160\arcdeg\ toward
due south \citep{vlba,2009AJ....138.1874L}, while the VLA data show
knots in a jet that curves from PA -170\arcdeg\
to the east between 5 and 10\arcsec\ from
the core.  In the X-ray image, the jet is detected most
strongly at a PA of -170\arcdeg\ out to 3.5\arcsec\ and is
marginally detected for much of the remainder of the jet. 
The readout streak is visible in the image shown
in Fig.~\ref{fig:images}.

\subsubsection*{2007$+$777 (S5 2007$+$77)}

This active galaxy is considered to be a BL Lac object with
hybrid FR I/II morphology \citep{2000A&A...363..507G}.
VLBI images show structure along a PA of -95\arcdeg\ with
a maximum apparent speed of 0.82$c$, the only jet in our
sample where the fastest components are clearly subluminal
\citep{lister01,vlba,2004A&A...428..847P}.
The VLA image shows a feature about 10\arcsec\ east of the core and a
20\arcsec\ long jet oriented at an average PA of -105\arcdeg\ with
some significant deviations from the average direction.
The {\em Chandra} image shows that the jet is detected along almost
its entire length and that the knot at the half-way point is particularly strong
\citep{2008ApJ...684..862S}.

\subsubsection*{2123$-$463 (PKS B2123$-$463)}

The southeast edge of a radio feature 4.5\arcsec\ from the core at a
PA of 110\arcdeg\ is
detected in X-rays.  There also appears to be some X-ray emission
about 1.5\arcsec\ from the core at a similar PA, associated
with a small-scale and bent radio jet.
The adopted redshift (1.67) may be very uncertain or even incorrect
\citep{2002A&A...386...97J}, being based on a unreliable objective prism spectrum.

\subsubsection*{2255$-$282 (PKS B2255$-$282)}

The 8\arcsec\ long radio jet extends along a PA of
 -70\arcdeg\ from the core.  The VLBI images show jet emission
 without detectable superluminal motion \citep{2007AJ....133.2357P}
 at a PA of -130 to -140\arcdeg (MOJAVE web site, \cite{2009AJ....137.3718L}).
 X-ray emission is detected along the first 4\arcsec\ of the
 jet and as close as 1\arcsec\ from the core
 (see Fig.~\ref{fig:xrayprofiles}).

\subsubsection*{2326$-$477 (PKS B2326$-$477)}

A strong radio component is detected about
4.5\arcsec\ from the core at a PA of -110\arcdeg.
VSOP imaging shows no jet at mas scales \citep{2004ApJS..155...33S}.
We detect no associated extended X-ray emission along this
PA.

\section{Discussion}

A hypothesis that bears testing with these data is that the
X-ray emission results from
IC-CMB photons off relativistic electrons and that the bulk motion
of the jet is highly relativistic and aligned close to the line of sight.
We have several lines of evidence that suggest that the jets in our sample
are consistent with this interpretation.

\subsection{Detection Statistics}

We detected 12 X-ray emitting jets among the 19 targets observed,
half of which were previously reported.
Of these detections, 9 were in the A subsample of 10 sources, while
only 3 were in just the B subsample: 0234$+$285, 2007$+$777, and 2123$-$463.
If detections were equally likely
in both B and A samples, then the {\it a priori} probability that there
would be $<4$ B detections would be 7.3\%, so the hypothesis that the
morphology selection is just as good as a flux selection is marginally
acceptable.
Of the aggregate of 39 sources from Paper I and this paper,
22 were in the A subsample.  Jets were detected in 16 of the 22 images,
for a 73\% detection rate.
This detection rate is similar
to that obtained by \citet{sambruna04} and \citet{marshall05}.
The jet detection rate for the B-only subsample
is not as high.  Of all the B-only quasars, 7 of 17 jets are detected (41\%).
These rates could be biased, however, because those targets observed in other
programs were generally the brightest A targets.

The typical X-ray flux densities of detected jets are greater than 2 nJy.
Flux densities in the radio band were generally lower in B targets than in
A targets while the X-ray flux limits are all about the same; consequently, the lower limits
on $\alpha_{rx}$ are higher for the A targets (see Fig.~\ref{fig:alpharx-z}).
However, Fig.~\ref{fig:arxdistribution} shows that the distribution of $\alpha_{rx}$ is
slightly shifted toward lower values of $\alpha_{rx}$ for B targets
compared to A targets, indicating
slightly larger X-ray flux densities relative to the jets' radio flux densities as a group.
Thus, it appears
that morphological selection may yield jets brighter in the X-ray band.
However, the distribution differences are not statistically significant, due
to small number of detected sources in the B subsample.  Furthermore, due to
the systematically higher redshifts of the B subsample, jet detection rate
differences between the two subsamples may result from
redshift dependences.

\subsection{Modeling the X-ray Emission}

\subsubsection{Distribution of $\alpha_{rx}$ and Redshift Dependence}

As in Paper I, there are bright X-ray jets even in sources with weak
extended radio flux, confirming that the ratio of the X-ray to
radio flux densities has a wide range (see Fig.~\ref{fig:arxdistribution}).
The $\pm 1\sigma$ width of the $\alpha_{rx}$ distribution
is about 0.15 -- a factor of 15 in $R$, the ratio of the X-ray and radio flux densities
(as extrapolated to a common frequency, see appendix~\ref{app:d95}).
The jets' radio flux densities extend over a factor of
almost 100 for the {\em detected} jets in our sample.  While there is a general
trend that the brightest jets are detected most often, it is clear that
predicting the X-ray flux from the radio knot flux densities is risky, so a shallow
survey is practically the only efficient means for finding jets that are X-ray bright.
We note that the two jets detected out of four exposures
longer than 10 ks would have been detected with just the first 10 ks.

Whereas detailed individual analyses of the brighter quasar jets 
can test physical models \citep[e.g.,][]{schwartz06,perlman08},
we explore here how even relatively
short exposures can prove useful for statistical tests of the
model in which the X-rays result from inverse Compton (IC)
scattering of cosmic microwave background (CMB) photons by
relativistic electrons in a jet moving with high bulk Lorentz
factor nearly along the line of sight (the IC-CMB model). Particular
support for this model arises in individual cases where
the optical flux lies below the radio to X-ray interpolation, 
indicating that synchrotron radiation
from a single population of relativistic electrons cannot fit the spectral energy
distribution.
We note that our objective is comparable to that examined by \citet{2004ApJ...600L..23C} and
\citet{ks05} but with a larger, more homogeneous sample of FR II quasars
containing a much larger
fraction of sources with $z > 1$.  We limit our analysis to the 34 quasars
in our sample
with known redshifts greater than 0.1 so as to avoid the slight contamination
by flat spectrum, core-dominated FR I radio galaxies.

Following \citet[][HK02]{hk02} and Paper I (see also Appendix~\ref{app:d95})
in the context of the IC-CMB model,
$R$ can be related to the
equipartition magnetic field in the absence of beaming, $B_1$,
derived from the radio flux and emitting volume, and beaming
parameters $\Gamma = (1-\beta^2)^{-1/2}$ and $\mu = \cos \theta$, where
$\theta$ is the angle to the line of sight, as
\begin{equation}
\label{eq:r}
R = A \bigg( \frac{b}{B_1} \frac{ (1 - \beta) (1+ \mu) }{(1-\mu \beta)^{2}}
	 \bigg)^{(1+\alpha)} (1+z)^{3+\alpha} 
\end{equation}

\noindent
where $A = 6.9 \times 10^{-21}$ and $b = 38080$~G are constants and $B_1$ has units of G.
The spectral index, $\alpha$, defined by $S_{\nu} \propto {\nu}^{-\alpha}$,
is assumed to be 0.8 for both the X-ray and radio bands.
Equation~\ref{eq:r} can be solved to give $\mu$ for an assumed value of $\beta$ (Paper I)
or for $\beta$ for a given value of $\mu$ \citep{2006xru..conf..643M, mc07}.
In Appendix~\ref{app:d95}, we show that the HK02 approach is equivalent to
the inverse Compton model developed by \cite{dermer95} which was later written in
a form independent of the system of units by \citet{2009A&ARv..17....1W}.

The quantity $R$ depends on
$\alpha_{rx}$ via the relationship
$R = (\nu_{\rm x}/\nu_{\rm r})^{\alpha-\alpha_{rx}}$,
so $\alpha_{rx}$ depends on quasar redshift in the IC-CMB model.
No significant
correlation of $\alpha_{rx}$ with $z$ is apparent in Fig.~\ref{fig:arxdistribution}.
However, with such a broad
distribution of $\alpha_{rx}$ it would be difficult to discern such a trend.
We tested the possibility that
$\alpha_{rx}$ depends on $z$ by splitting the sample into two redshift ranges.
For $0.55 < z < 0.95$, the average $\alpha_{rx}$
is 1.001 $\pm$ 0.020, compared to a value of 0.954 $\pm$ 0.019 for
$0.95 < z < 2$.   The difference is insignificant.

A more sensitive test is to explore the dependence of $R$ upon $z$.
We use the method developed by \cite{marshall92} to fit
$R$ to the form $(1+z)^a$; details are given in Appendix~\ref{app:regress}. 
However, $B_1$ is calculated from observations and depends on
redshift according to model assumptions (as discussed
below).  Generally, we expect $B_1 \propto f(z)$, giving

\begin{equation}
\label{eq:rz}
R  \propto (1+z)^{3+\alpha} [f(z)]^{-(1+\alpha)}
\end{equation}

\noindent
In the simple case where the distribution of intrinsic magnetic fields
is independent of redshift, then we may set $f(z) = 1$.  The log likelihood for
this case is shown in
Fig.~\ref{fig:arxtest}, for which we find 
$a = 0.7 \pm 1.6$ at 90\% confidence
($\Delta \chi^2 = 2.71$).  The likelihood ratio test then rejects $a >
3.5$ at $>99$\% confidence, whereas we expect $a = 3.8$ for
the IC-CMB model.

It is common to estimate the magnetic field in individual
sources based on
observations and assume minimum energy.
We note that a simple dependence of $B_1$ calculated this
way (Tables~\ref{tab:knotresults} and \ref{tab:beaming})
with $z$ is not readily apparent in
our data (see Fig.~\ref{fig:b1vz}), but other factors entering the calculation of $B_1$
(particularly, the jet's radio flux density and angular length)
have a broad scatter and probably serve to mask any relationship.

A simple case to consider is one described by
\citet{2009A&ARv..17....1W}.  If source volume is estimated via
angular
sizes in two dimensions assuming that the path through the jet is
independent of redshift, then 
the volume $V \propto d_{\rm A}^2 \propto d_{\rm L}^2/(1+z)^4$, where
$d_{\rm A}$ and $d_{\rm L}$ are angular and
luminosity distances, respectively.\footnote{The values of the volume
reported in Paper I were computed incorrectly, so we provide the
correct values of $V$, $B_1$, $K$ in Table~\ref{tab:beaming}.
The sense of the error is that the volumes in Paper I were too large,
causing $B_1$ and $K$ to be too small by about a factor of 10
in some cases, and $\theta$ to be
about a factor of 2 larger than we now find.}
For minimum energy (or equipartition),

\begin{equation}
\label{eq:fdmw}
f(z) \propto  \bigg [ \frac {L_{\rm s}(z)}{V(z)} \bigg ]^{1/(\alpha+3)}
\propto  \bigg [ \frac {(1+z)^{(\alpha -1)} d_{\rm L}^2}{V(z)} \bigg
]^{1/(\alpha+3)}
\propto (1+z)
\end{equation}

\noindent
Here we have assumed that the minimum-energy field is measured
over fixed electron energies in the rest frame of the source.
In the case of calculations over fixed frequencies in
the observer's frame ($10^7$ to $10^{15}$ Hz are actually adopted for
$B_1$ in Tables 5 and 6) the result is similar, with
exponent $2/7$ rather than
$1/(\alpha+3)$, where (as in Paper I) we assume that
$\alpha=0.8$.
Combining Equations \ref{eq:rz} and \ref{eq:fdmw} gives 

\begin{equation}
\label{eq:Rdmw}
R \propto (1 + z)^2
\end{equation}

\noindent
which agrees with equation 13 of \citet{2009A&ARv..17....1W}.
Under this assumption for the jet volume, the fit value of $a$ is consistent
with the prediction of the IC-CMB model.

Alternatively, the volume can be estimated
assuming that the jet is
a cylinder of constant angular radius matched to {\em Chandra}'s resolution
(as adopted in Paper I and used for the estimates of $B_1$ in this
paper).  Here, $V(z) \propto d_{\rm A}^3$, so $d_{\rm A}$ does not cancel
in the equations, giving

\begin{equation}
\label{eq:fmh}
f(z) \propto (1+z) d_A^{-{1/(\alpha + 3)}}
\end{equation}

\noindent
In this case, $f(z)$ does not have a simple dependence on
$(1+ z)$ over the redshift distribution of our sources.  Instead, we
define a new quantity that is derived from the observed data for each source,
$Q \equiv R B_1^{1+\alpha}$.  In the IC-CMB model,
$Q \propto (1+z)^{3+\alpha}$, while our fit to $Q \propto (1+z)^{a}$
gives $a = 1.35 \pm 1.36$ (at 90\% confidence).
Here $a = 3+\alpha$ is rejected at better than 99\% confidence
for $\alpha > 0.5$.
The best fit resulted in a smaller index,  $a = -0.37 \pm 1.35$,
and $a = 3.5$ is still rejected at 99\% confidence.

Thus, we have two circumstances where the
IC-CMB model can be ruled out and one in which
it is still viable, where the jet volume is computing using the
assumption described above Equation \ref{eq:Rdmw}.
The circumstances involve different but plausible conditions
dictating the dependence of the intrinsic magnetic field with
redshift, so it is difficult to provide a definitive test using these data alone.
The factors that go into estimating the magnetic field
bear further investigation as source details are obtained in
follow-up radio, X-ray, and optical observations in order to develop a refined
test of the model.  One source of uncertainty in our method
of using the X-ray and radio emission for the entire jet rather
than for individual knots is that the jet geometries are often
complex.  Furthermore, the termination knots may also be
included in some cases, where it is unlikely that both the
radio and X-ray emission regions are moving relativistically
relative to the nucleus.  This paper is concerned primarily with
shallow observations and deeper individual analyses would
be best suited to examine these more subtle issues.

\subsubsection{Angles to the Line of Sight}

As in Paper I, we computed the distribution of angles to
the line of sight for these kpc scale jets,
under the assumptions that 1) X-rays arise from the
IC-CMB mechanism, and 2) all jets have a common
Lorentz factor, $\Gamma$.
\cite{vlba} estimated the intrinsic Lorentz factor distribution for a flux-selected
set of core-dominated quasars, finding that it appears broad, with most
values of $\Gamma$ between 5 and 25 (see their Fig.~9).
For now, we assume $\Gamma = 15$ and find that $\theta$ ranges from
6\arcdeg\ to 13\arcdeg\ for the quasars in our sample
(see Table~\ref{tab:beaming}).  For these sources, the Doppler factor, $\delta$,
is in the range 3-8, compared to the assumed Lorentz factor of 15. 

Because the jet surface brightness is not constant and the spatial
variations between the radio and X-ray bands can differ, it is possible
that systematic errors result from considering the entire jet.
To estimate the effect of restricting attention to knots within the jets,
we have computed X-ray and radio flux densities
for a selection of 3\arcsec$\times$3\arcsec\ regions from the jets.
Measurements are given in Table~\ref{tab:knotresults} and angles to the
line of sight are given in Table~\ref{tab:knotbeaming}.  The angles usually decrease
by a degree or less from the full-jet estimates.
For the remainder of this section, we will only consider
results for the entire jet, leaving analysis of individual knots to follow-up work which will
require deeper X-ray observations with higher knot counts.
See section~\ref{sec:sources} for comments about individual sources
and references to more detailed analyses, where available.

Many of these sources are in the MOJAVE program, which consists of
VLBI observations of several hundred compact active galaxies and
quasars used to measure pc-scale proper motions.
Of the 22 sources in common with our sample, we have X-ray data for 14,
as listed in Table~\ref{tab:bends}.  For all but one quasar of the 22, there is apparent
superluminal motion.  Values of $\beta_{\rm app}$, the apparent velocity of the
most rapidly moving pc-scale component relative to $c$, are given in the table.
See the discussion of individual sources for references.

The population modeling by \citet{2007ApJ...658..232C} based on the
MOJAVE sample provides a
basis for testing the IC-CMB model for our sample.  As a first step, it is important
to determine that our sample is a representative subset of the MOJAVE sample.
For the flux-limited MOJAVE sample, \citet{2007ApJ...658..232C} showed that $\sin \theta$
of the pc-scale jets are generally within 50\% of $1/\beta_{\rm app}$.
Fig.~\ref{fig:sldistribution} shows that the distribution of $1/\beta_{\rm app}$
for our sample is as concentrated below about 5\arcdeg\ as the MOJAVE sample.
Also shown in this figure is the distribution of the values of $\theta$ for the large
scale jets, as derived from the IC-CMB model.  These angles are generally below
11\arcdeg\ but systematically larger than the angles estimated for the pc-scale
jets.  This difference is not surprising because the pc-scale and kpc-scale
jets are not aligned in projection on the sky but suggests that most misalignments
are small.  Position angle differences are given
in Table~\ref{tab:bends}.

We now attempt to quantify the comparison of the angles to the line of sight for
the kpc-scale jets with information in the pc-scale jets.  In appendix~\ref{app:angle},
we show how one may estimate the range of kpc-scale angles to the line of sight
by using only the values of $\beta_{\rm app}$ for the pc-scale jets and the
position angle differences.  At the same time, intrinsic bend angles, $\zeta$, are
estimated and a probable range for these angles are computed.  The
 results are given in Table~\ref{tab:bends} and shown in
Fig.~\ref{fig:angles}, where it can be seen that these
independent estimates are generally consistent.  However, there are
some notable exceptions, particularly where the angles from the
IC-CMB calculation are of order a factor of two larger than those
based on geometry and superluminal motion of the pc-scale jet.  For
these exceptions, one may infer that the jets decelerate
substantially from pc to kpc scales.

\section{Summary}

We have reported new imaging results using the {\em Chandra} X-ray Observatory
for quasar jets selected from the sample originally defined by \citet[][Paper I]{marshall05}.
For the larger sample, we confirm many results in Paper I: 1) quasar jets can be readily detected
in X-rays using short {\em Chandra} observations, 2) no X-ray counterjets are detected,
and 3) the line-of-sight angles of the kpc-scale jets are small in the IC-CMB model
in which the X-ray emission results from inverse Compton upscattering of cosmic
microwave background photons by relativistic electrons in a jet moving with large
bulk Lorentz factor.

In addition, we have several new results: 1) depending on how the jet volume
is computed, which determines how estimates of the intrinsic magnetic
field may vary with $z$, $S_x/S_r$ values do not increase with
$1+z$ at the level expected in a simple IC-CMB model, and
2) the pc-scale jet speed and orientation indicate
that some kpc-scale jets are oriented closer to the line of sight than inferred from the
IC-CMB model, suggesting instead that these jets decelerate from pc to kpc scales.
Other results also cast doubt upon a pure IC-CMB interpretation
of the X-ray emission from kpc-scale jets \citep{2006ApJ...648..910U,2006ApJ...648..900J}, tending
toward an interpretation requiring a high energy population of electrons producing
synchrotron emission observable in the optical to X-ray bands but undetected in
the radio band.  While  we may not be able to verify this interpretation using these
short X-ray exposures, our results do indicate that the large scale jets are likely
to be seen in projection at small angles to the line of sight, as assumed in the
IC-CMB model.  In this case, in order to {\it avoid} substantial inverse Compton
emission, the kpc-scale jets must have smaller Lorentz factors than their
pc-scale counterparts.

\appendix

\section{Regression without Limits}

\label{app:regress}

Here we give the formal method for handling a linear regression where some of the
data points have such large error bars that a logarithmic transformation is
mathematically disallowed.  One approach for handling such data is to degrade
the poor measurements by assigning limits to them and then using a regression
method developed by \citet{at86}.  An alternative is to use the error bars and
distinguish between the observation and the intrinsic quantities.  A regression
method using all error bar information was presented by \cite{marshall92} that
we shall use here.  See that paper for a discussion of the approach.

Our objective is to determine $a$ in the model

\begin{equation}
\log R = a \log x + b
\end{equation}

\noindent
where $x \equiv 1+z$.  The data set is ${R_n, \sigma_n, x_n}$, where $R_n$ are
the estimated luminosity ratios that have uncertainties $\sigma_n$ that may be
large.  Note that some $R_n$ values may be negative due to large background
and low intrinsic $R$ values.  Following \cite{marshall92}, we define the probability
model for the true $R$, when given $x$ and the model parameters as

\begin{equation}
p(R; x; a, b, s) = \frac{1}{(2 \pi)^{1/2}s} \exp (-\frac{( \log R - a \log x - b)^2 }{2s^2} )
\end{equation}

\noindent
where $s$ is the standard deviation of the population about the trend line we
are fitting.  The likelihood for the observed values, given this model and the
(Gaussian) uncertainties on the measurements is

\begin{equation}
L = \prod_n p\prime(R_n; x_n, \sigma_n; a, b, s) = \prod_n
 \int_0^\infty p(R; x_n; a, b, s) \frac{1}{(2 \pi)^{1/2} \sigma_n} e^{\frac{ -(R_n - R)^2}{2\sigma_n^2} } d R
\end{equation}

\noindent
so that the minimization statistic that is distributed as $\chi^2$ is $-2 \log L$.  The integration
interval is set by the physical condition that $R$ cannot truly be less than zero.

\section{The Dependence of $R$ upon Beaming Parameters}

\label{app:d95}

We have used the formalism of \citet[][HK02]{hk02} in equation~\ref{eq:r}.
We could just as well have used the derivation by \citet[][D95]{dermer95}.
We now show that these two approaches yield the same dependence of $R$
upon $\mu$, $\beta$, and $z$.

Much of the reconciliation comes in the definition of terms.  Translating
from D95's notation to HK02 terms,
$\beta_\Gamma = v/c$ becomes $\beta$, $\mu_e = \mu$ is the cosine of the emission angle
in the rest frame of the relativistically moving jet knot
and corresponds to $\mu_j^\prime$, while $\mu_{\rm obs}$ is the cosine of the angle of
the jet to the observer's line of sight and corresponds to $\mu$.  The identity
cited by D95 just before equation 7 is

\begin{equation}
\Gamma (1+\mu) = D (1+\mu_{\rm obs})/(1 + \beta_\Gamma)
\end{equation}

\noindent
corresponds to the readily proved identity in HK02 notation

\begin{equation}
\label{eq:gmu}
\Gamma (1+\mu_j^\prime) = \delta (1+\mu)/(1 + \beta)
\end{equation}

\noindent
where $D$ (in D95 notation) is $\delta = \frac{1}{\Gamma (1- \beta \mu)}$, the Doppler boost factor
in HK02 notation.

D95's equations 7 and 8 define the IC-CMB flux, $S_C$, and the synchrotron
flux, $S_{\rm syn}$, so taking the ratio gives

\begin{equation}
\label{eq:d95}
R \equiv \frac{S_{\rm x} (\nu/\nu_{\rm x})^{-\alpha}}{S_{\rm r} (\nu/\nu_{\rm r})^{-\alpha}} =
	S_C / S_{\rm syn} = 
	\frac{3 u_{\rm iso}^*}{2 u_B} D^{1+\alpha}
	\bigg [ \frac{1+\mu_{\rm obs}}{1+\beta_\Gamma} \bigg ]^{1+\alpha} (\epsilon_B/\epsilon^*)^{1-\alpha}
\end{equation}

\noindent
in D95 notation, where $u_{\rm iso}^*$ is the monochromatic, isotropic, radiation density in the
host galaxy's rest frame;
$u_B$ is the magnetic field energy density in the knot rest frame;
$\epsilon_B \equiv B/B_Q$ for $B_Q = 4.414 \times 10^{13}$ G;
and $\epsilon^*$ is a dimensionless form of the photon energy of the isotropic radiation in the host rest frame.
In general, $u_{\rm iso}^* =
u_{\rm iso,0}^* (1+z)^4$ and $\epsilon^* = \epsilon_0^* (1+z)$, where the subscript 0 indicates
the corresponding quantities at the current epoch ($z = 0$).
Equation~\ref{eq:d95} is then the same as equation 7 of \citet{2009A&ARv..17....1W} who expresses
it in a form fully independent of the system of units by
using the gyrofrequency, $\nu_{\rm g}$, in place of $\epsilon_B$.

When $B$ is estimated using minimum energy arguments, then $B = B_1 / \delta$ (HK02).
Substituting gives

\begin{equation}
R = \frac{C_0}{B_1^{1+\alpha}}
	\bigg [ \frac{\delta^2 (1+\mu)}{1+\beta} \bigg ]^{1+\alpha} (1+z)^{3+\alpha}
\end{equation}

\noindent
in HK02 notation, where $C_0 = 6 \pi u_{\rm iso,0}^* (B_Q \epsilon_0^*)^{\alpha-1}$.

Substituting the identity from equation~\ref{eq:gmu} and rearranging gives

\begin{equation}
\Gamma \delta (1+\mu_j^\prime) \propto B_1 (R/R_0)^{1/(1+\alpha)} (1+z)^{-(3+\alpha)/(1+\alpha)}
\end{equation}

\noindent
which has identical dependences on $z$, $\beta$, and $\mu$ as derived by HK02.
The dependence on beaming parameters is also the same as given by \citet[][Eq. 13]{2009A&ARv..17....1W}.
\citet{2009A&ARv..17....1W} also pointed out that $B_1$ depends
on $z$ in a model-dependent way, so the dependence on $z$ is not the same as derived here.

\section{Limiting Angles to the Line of Sight}

\label{app:angle}

We follow and extend the analysis of \citet{1993ApJ...411...89C}, where they estimate jet
bend angles.  In their notation, $\theta$ is the angle of the jet to the line of sight
before a bend, $\zeta$ is the magnitude of the jet bend, $\phi$ is a phase angle
giving the rotation of the bent jet about the axis defined by the jet before the
bend, and $\eta$ is the apparent bend, as projected on the sky.  For this
definition,

\begin{equation}
\label{eq:eta}
\tan \eta = \frac{\sin \zeta \sin \phi}{\cos \zeta \sin \theta + \sin \zeta \cos \theta \cos \phi}  .
\end{equation}

\noindent
We may solve this equation for $\zeta$:

\begin{equation}
\label{eq:zeta}
\tan \zeta = \frac{\tan \eta \sin \theta}{\tan \eta \cos \phi \cos \theta - \sin \phi }  .
\end{equation}

\noindent
On the right hand side is the observable, $\eta$, the angle to the line of sight that
may be estimated using $\sin \theta = 0.5/\beta_{\rm app}$
\citep{1994ApJ...430..467V,1997ApJ...476..572L}, and the
unknown phase angle, $\phi$.
Taking a somewhat Bayesian view, we assume that this angle is uniformly
distributed between $0$ and $2 \pi$, so that one may assign a probability to
possible values of $\zeta$.  For small $\theta$ and $\eta$, $\zeta$ is also small
with high probability; in essence, it is unlikely that the intrinsic jet bends are large
if the pc-scale jet is nearly aligned to the line of sight and the position angle
difference is relatively small.  In practice, rather than setting $p(\theta) = \frac{1}{2 \pi}$
and solving for $p(\zeta)$, we tabulate $\zeta(\phi; \theta, \eta)$, for each source
and determine the points at which $p(>\zeta)$ equals 1, 0.5, or 0.10 to give the
minimum, mid-range, and maximum values of $\zeta$.

We extend this analysis by defining the angle of the jet to the line of sight after
the bend to be $\xi$.  We find

\begin{equation}
\cos \xi = \cos \zeta \cos \theta - \sin \zeta \cos \phi \sin \theta
\end{equation}

\noindent
so that, when given a value of $\zeta$, one may determine $\xi$.  We follow
a similar procedure for determining the
minimum, mid-range, and maximum values of $\xi$ as was followed for
determining the range of possible values for $\zeta$.

\acknowledgments

We thank Marshall Cohen for communicating results in advance of
publication.
Support for this work was provided in part by the National Aeronautics and
Space Administration (NASA) through the Smithsonian Astrophysical Observatory (SAO)
contract SV3-73016 to MIT for support of the Chandra X-Ray Center (CXC),
which is operated by SAO for and on behalf of NASA under contract NAS8-03060.
Support was also provided by NASA under contract NAS 8-39073 to SAO.
JMG was partially supported under Chandra grant GO4-5124X to MIT from the CXC.
This research has made use of data from the MOJAVE
database that is maintained by the MOJAVE team \citep{2009AJ....138.1874L}.
This research has made
use of the United States Naval Observatory (USNO) Radio Reference
Frame Image Database (RRFID).  The Australia Telescope Compact
Array is part of the Australia Telescope which is funded by the
Commonwealth of Australia for operation as a National Facility managed
by CSIRO.  This research has made use of the NASA/IPAC Extragalactic
Database (NED) which is operated by the Jet Propulsion Laboratory,
California Institute of Technology, under contract with the
National Aeronautics and Space Administration.

{\it Facilities:} \facility{CXO(ACIS)}, \facility{ATCA}, \facility{VLA}

\clearpage

\begin{figure}[htp]
  \epsscale{0.8}
  \plotone{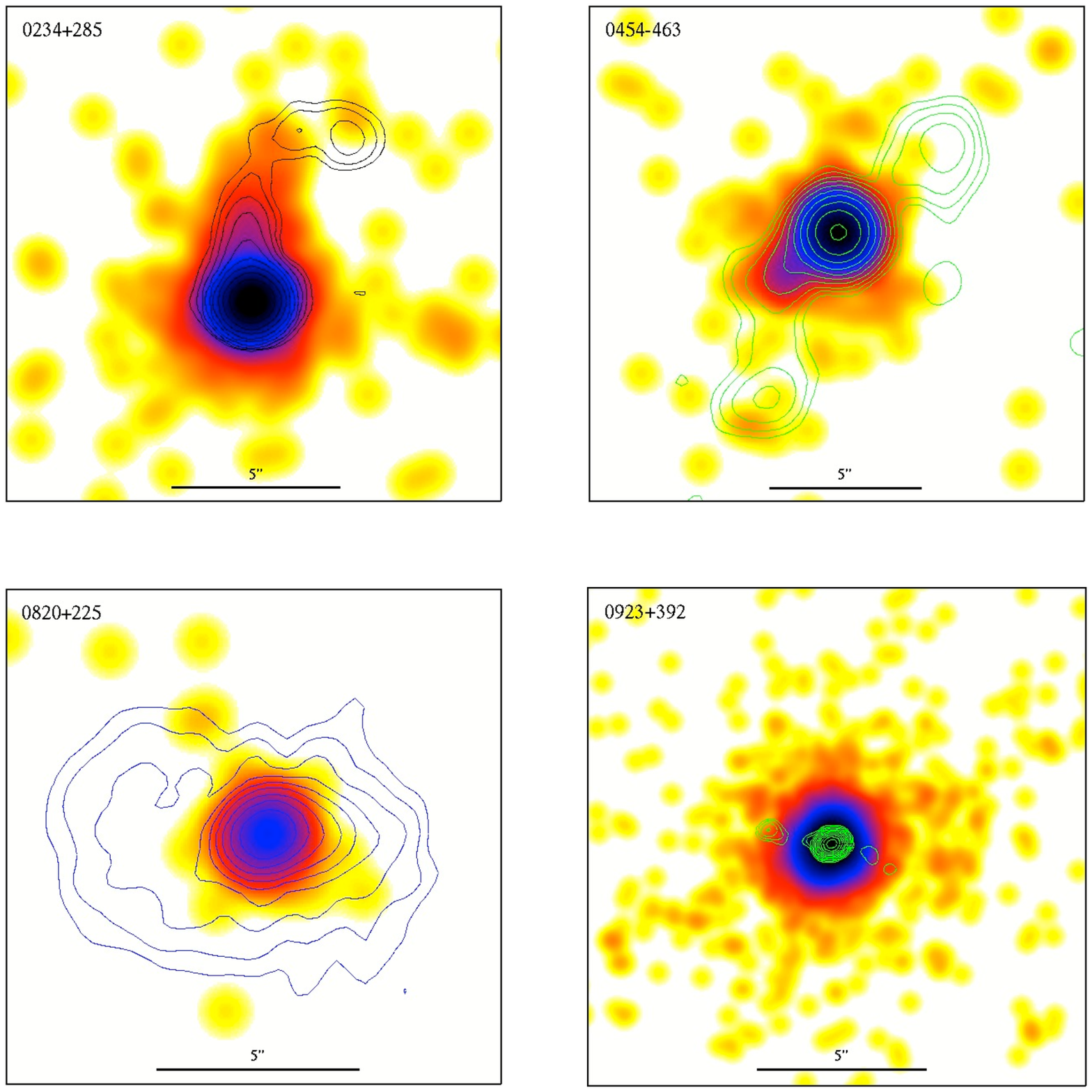}
  \caption{X-ray images obtained with the {\em Chandra} X-ray Observatory,
  overlaid by contours of radio emission obtained at the Australia
  Telescope Compact Array or the Very Large Array (VLA).
  The images appear in the following order:
  {\em a)} 	0234$+$285,
  {\em b)} 	0454$-$463,
  {\em c)} 	0820$+$225,
  {\em d)} 	0923$+$392,
  {\em e)} 	0954$+$556,
  {\em f)} 	1040$+$123,
  {\em g)} 	1055$+$018,
  {\em h)} 	1055$+$201,
  {\em i)} 	1116$-$462,
  {\em j)} 	1251$-$713,
  {\em k)} 	1354$+$195,
  {\em l)} 	1421$-$490,
  {\em m)} 	1641$+$399,
  {\em n)} 	1642$+$690,
  {\em o)} 	1928$+$738,
  {\em p)} 	2007$+$777,
  {\em q)} 	2123$-$463,
  {\em r)} 	2255$-$282, and
  {\em s)} 	2326$-$477. 
  The radio surface brightnesses increase by $\times 2$ for
  each radio contour, starting at five times the rms noise (from Table~\ref{tab:radioCont}).
  The X-ray images are convolved with 1\arcsec\ Gaussians and then
  binned at 0.0492\arcsec.  The color scales are the
  same in all images, ranging logarithmically
  from 0.5 counts/beam (yellow) to 2500 counts/beam
  (black).  Notes on individual objects are given in the text.
  For sources with bright cores, a readout streak may be observed
  on both sides of the core, such as in 1055$+$018.
  } \label{fig:images}
\end{figure}

\addtocounter{figure}{-1}

\begin{figure}[htp]
\epsscale{1.0}
  \plotone{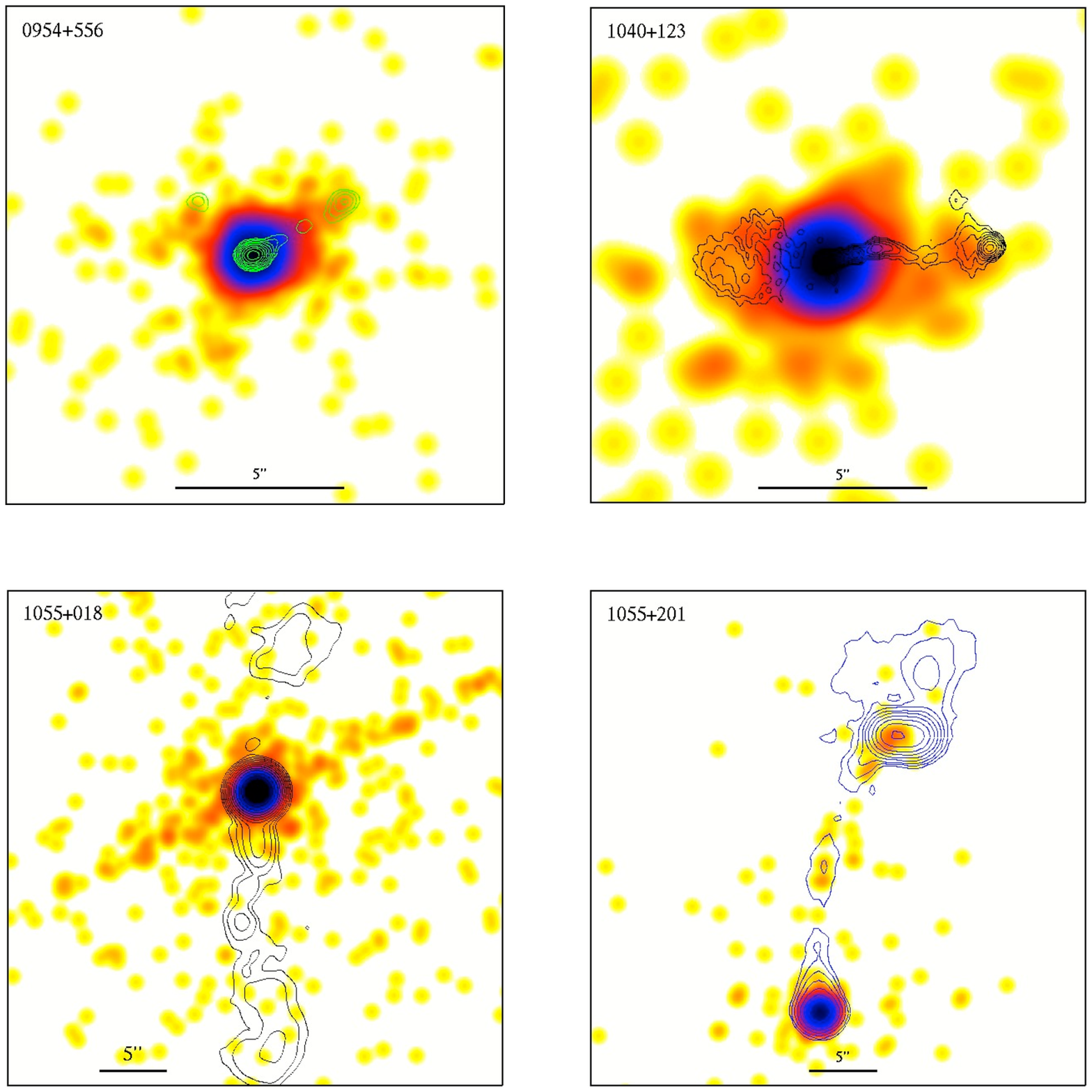}
  \caption{continued.}
\end{figure}

\addtocounter{figure}{-1}

\begin{figure}[htp]
  \plotone{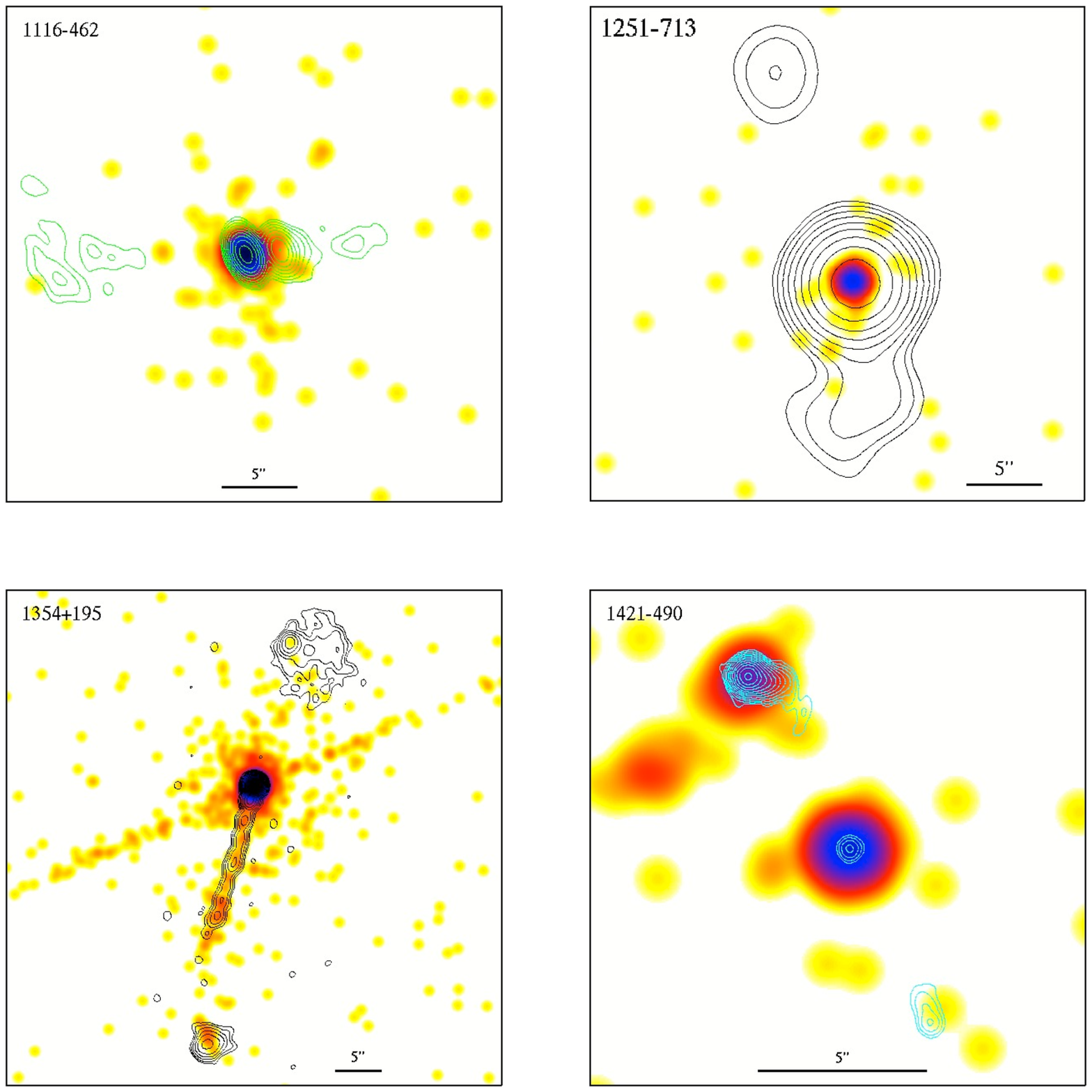}
  \caption{continued.}
\end{figure}

\addtocounter{figure}{-1}

\begin{figure}[htp]
  \plotone{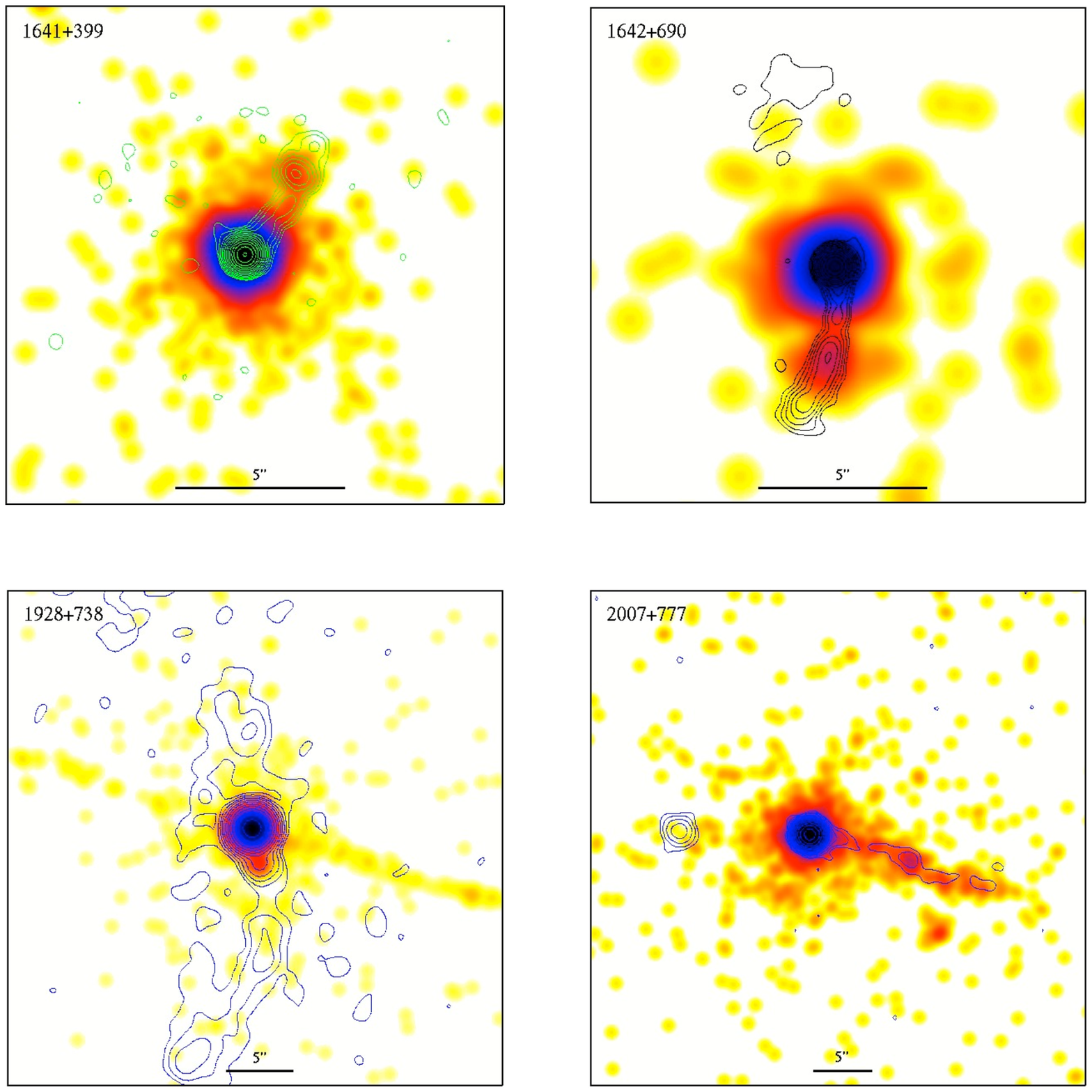}
  \caption{continued.}
\end{figure}

\addtocounter{figure}{-1}

\begin{figure}[htp]
  \plotone{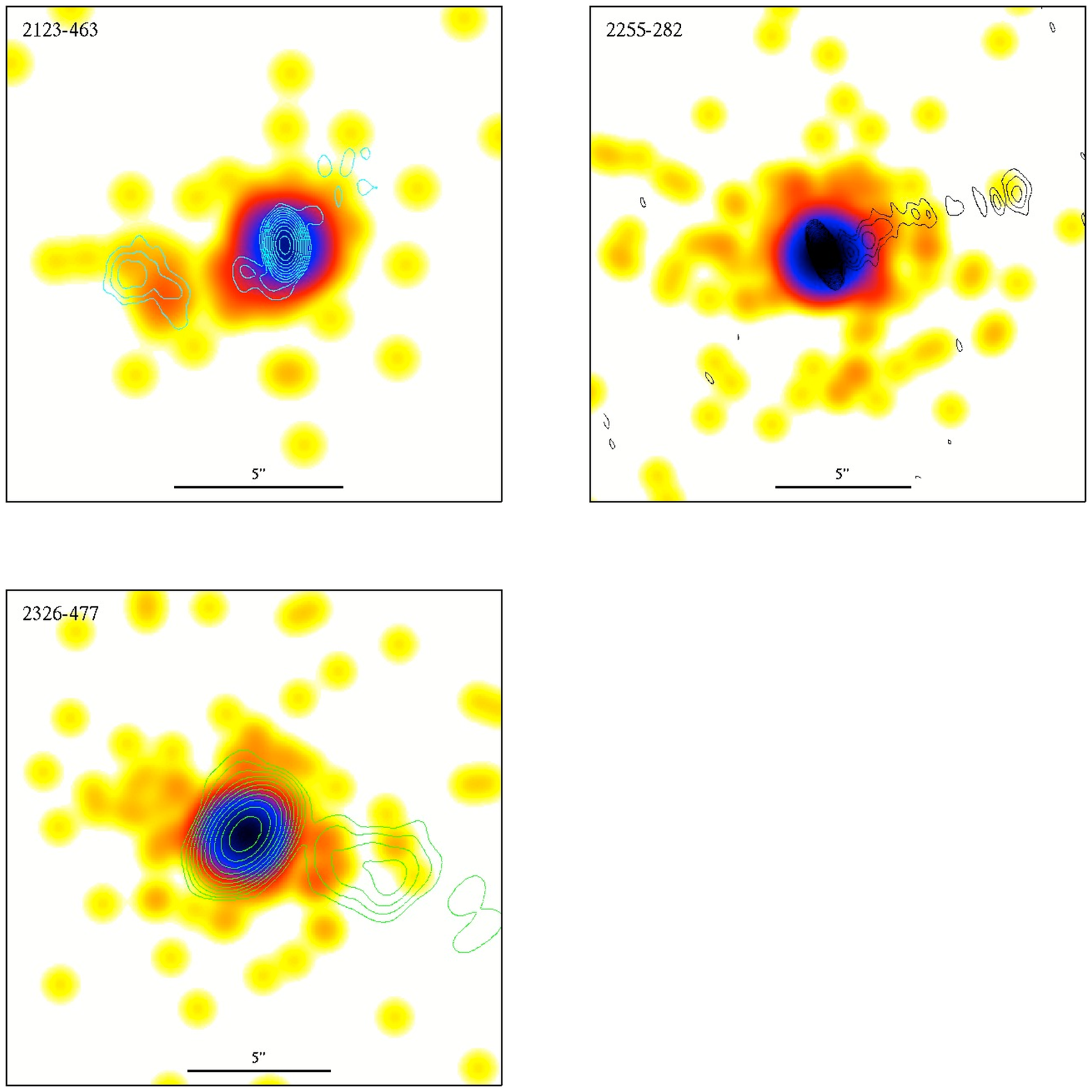}
  \caption{continued.}
\end{figure}

\begin{figure}
\epsscale{0.8}
\plotone{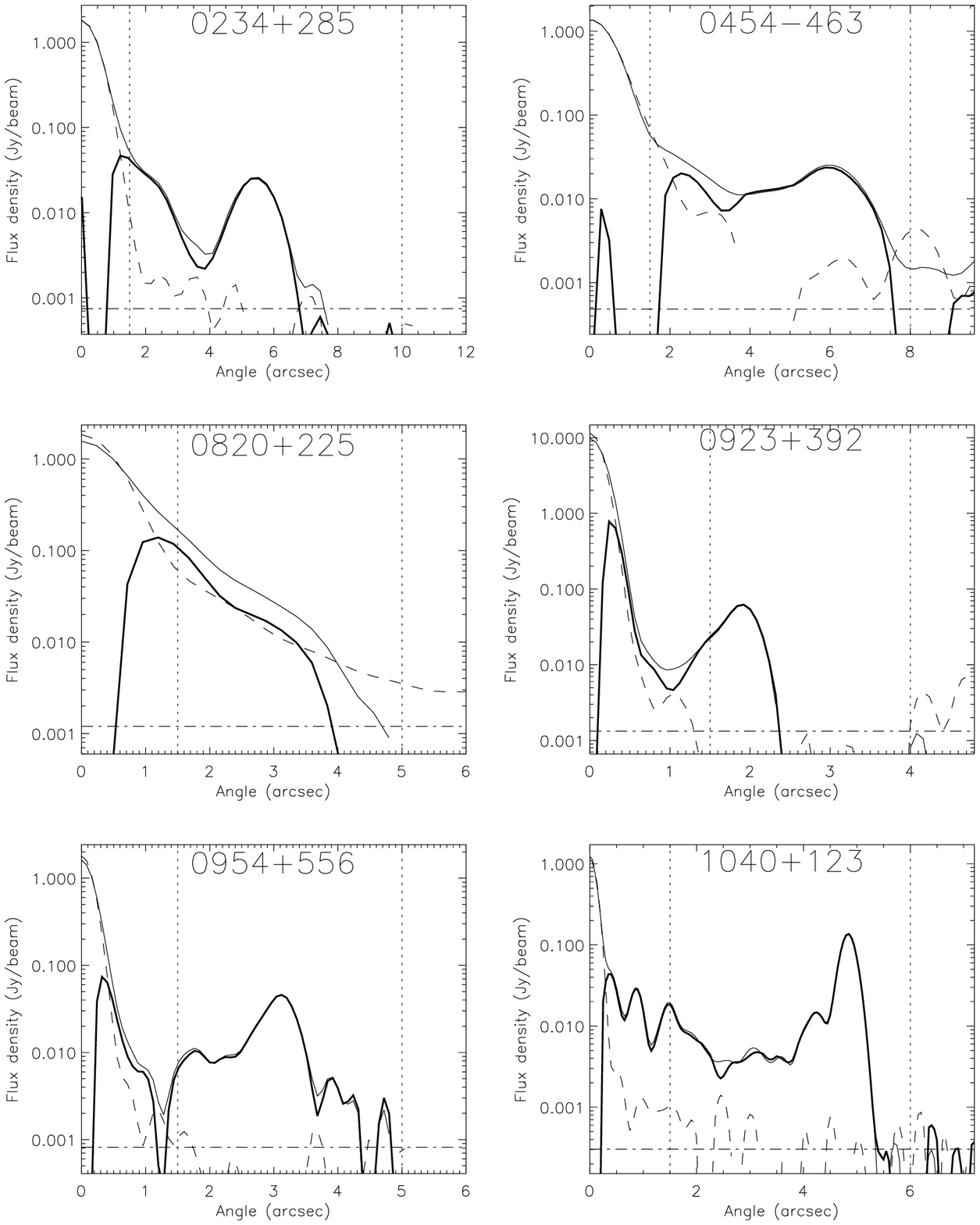}
  \caption{Profiles of the radio images.  The solid, thin
  lines give the profiles along the position angles of the jets, as defined
  in Table~\ref{tab:jetresults} and used for measuring the X-ray profiles.
  The integrated flux densities are determined in rectangles given by
  parameters listed in Table~\ref{tab:jetresults}.
  The dashed lines give the profiles at
  a position angle 90\arcdeg\ CW from the jet to avoid any counter-jets
  or lobes opposite the jet.
  The solid, bold lines give the difference between the profiles along the
  jet and perpendicular to it, so that the core is effectively nulled and
  the jet flux can be measured as a residual between the vertical dotted lines.
  The horizontal dash-dot lines give the average noise level in the map.
  } \label{fig:radioprofiles}
\end{figure}

\addtocounter{figure}{-1}

\epsscale{1.0}
\begin{figure}
\plotone{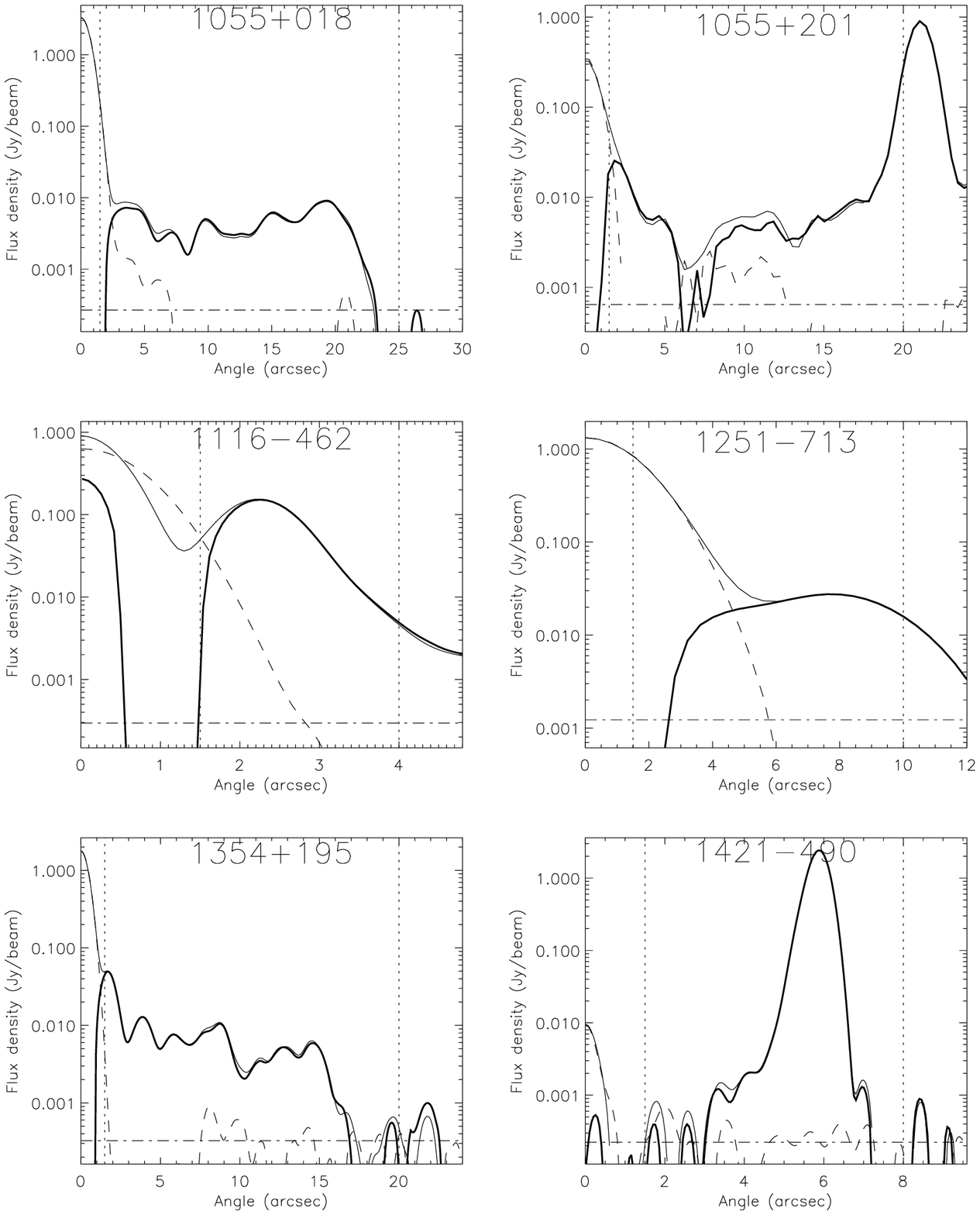}
  \caption{continued.}
\end{figure}

\addtocounter{figure}{-1}

\begin{figure}
\plotone{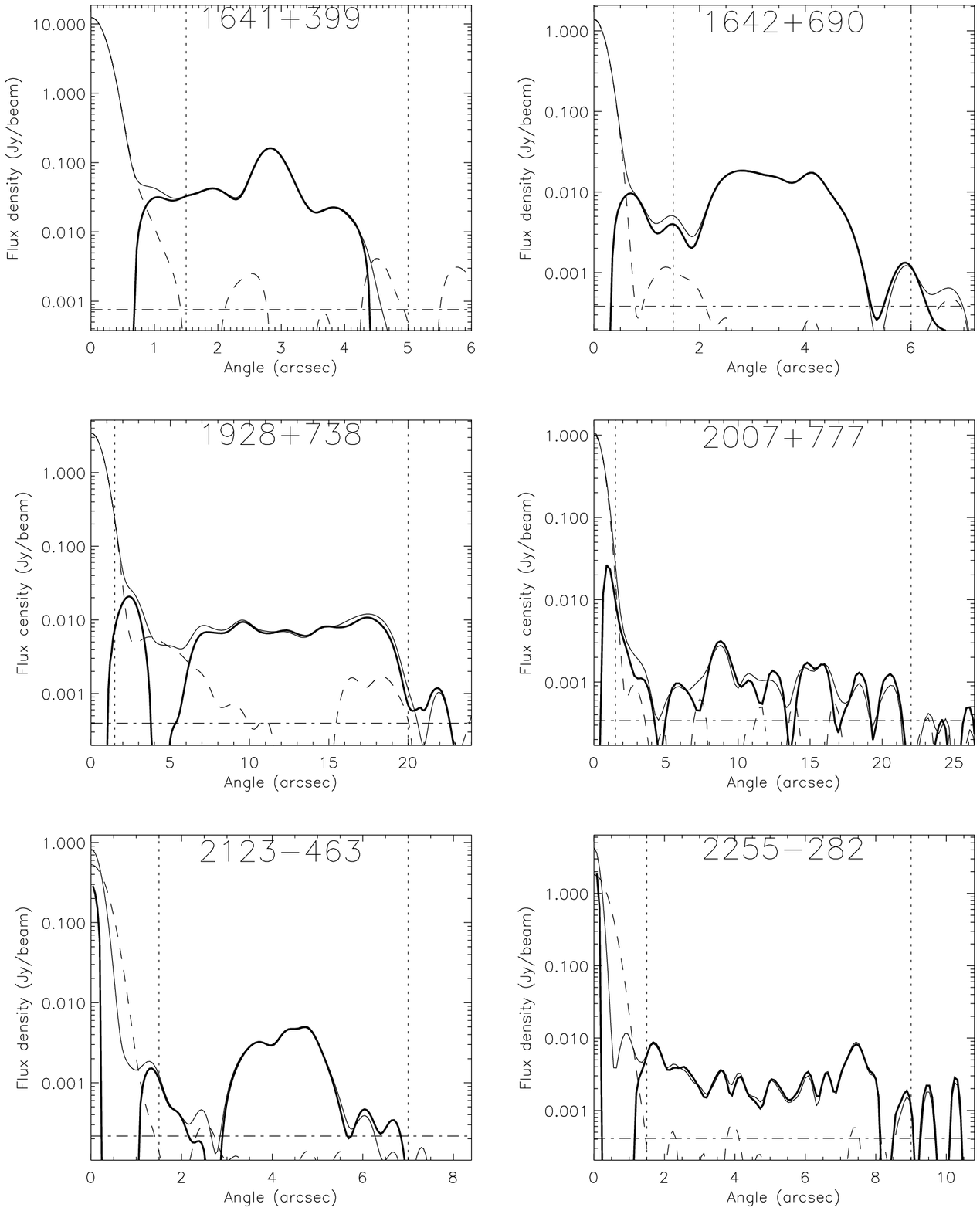}
  \caption{continued.}
\end{figure}

\addtocounter{figure}{-1}

\begin{figure}
\plotone{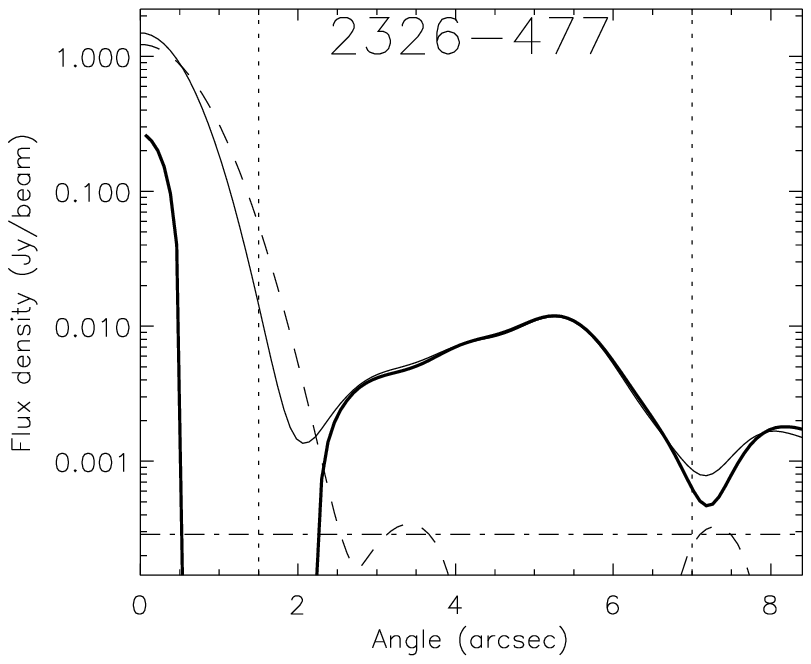}
  \caption{continued.}
\end{figure}

\begin{figure}
\plotone{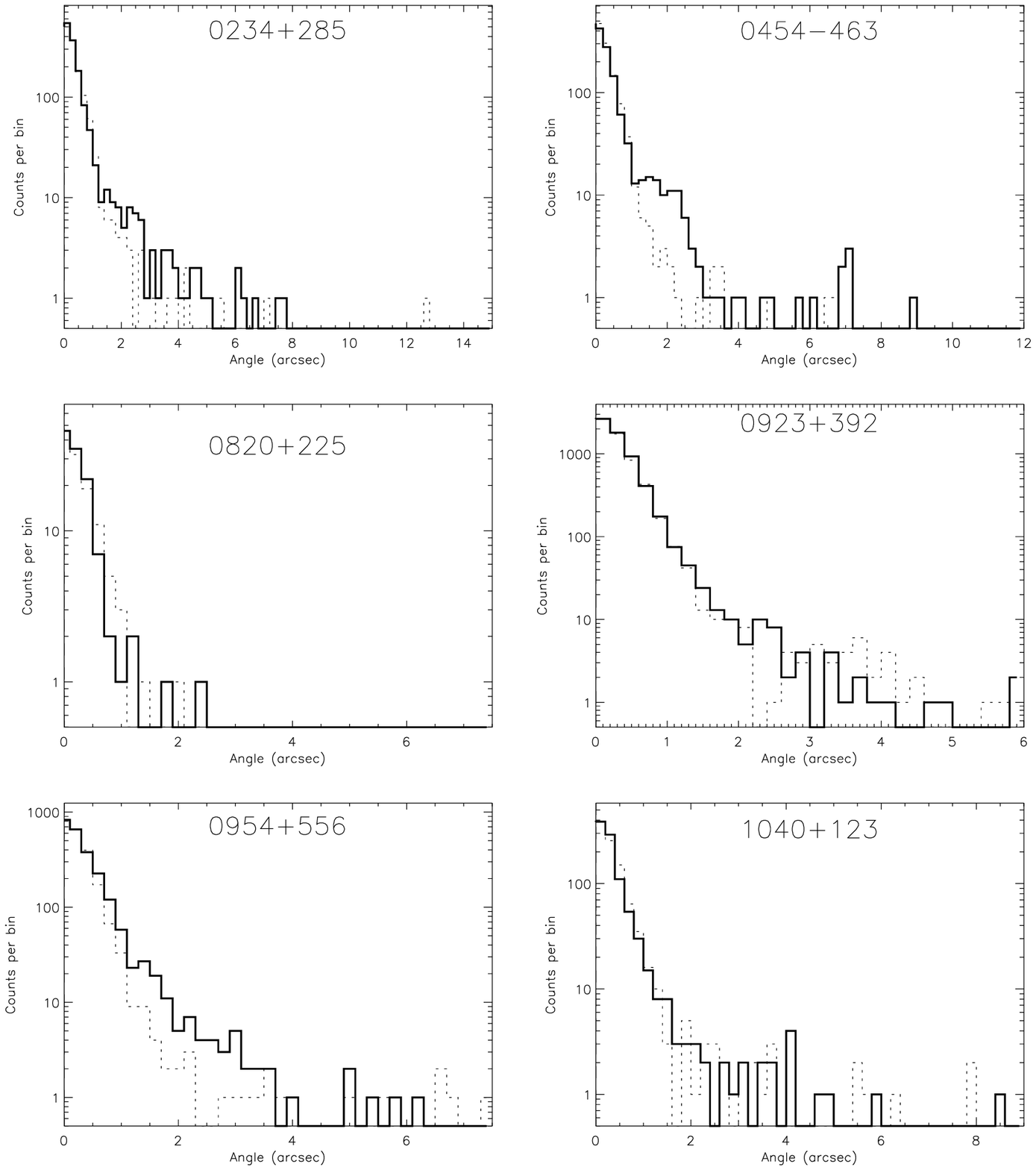}
  \caption{Histograms of counts from the X-ray images.  The solid, bold
  lines give the profiles along the position angles of the jets, as defined
  in Table~\ref{tab:jetresults} and used in Fig.~\ref{fig:radioprofiles}.
  The dashed histograms give the profiles at
  a position angle 180\arcdeg\ opposite to the jet -- the counter-jet
  direction.  The counter-jet profiles provide a measure of the significance
  of the X-ray emission from the jet because there are no clearly detected
  counter-jets.} \label{fig:xrayprofiles}
\end{figure}

\addtocounter{figure}{-1}

\begin{figure}
\plotone{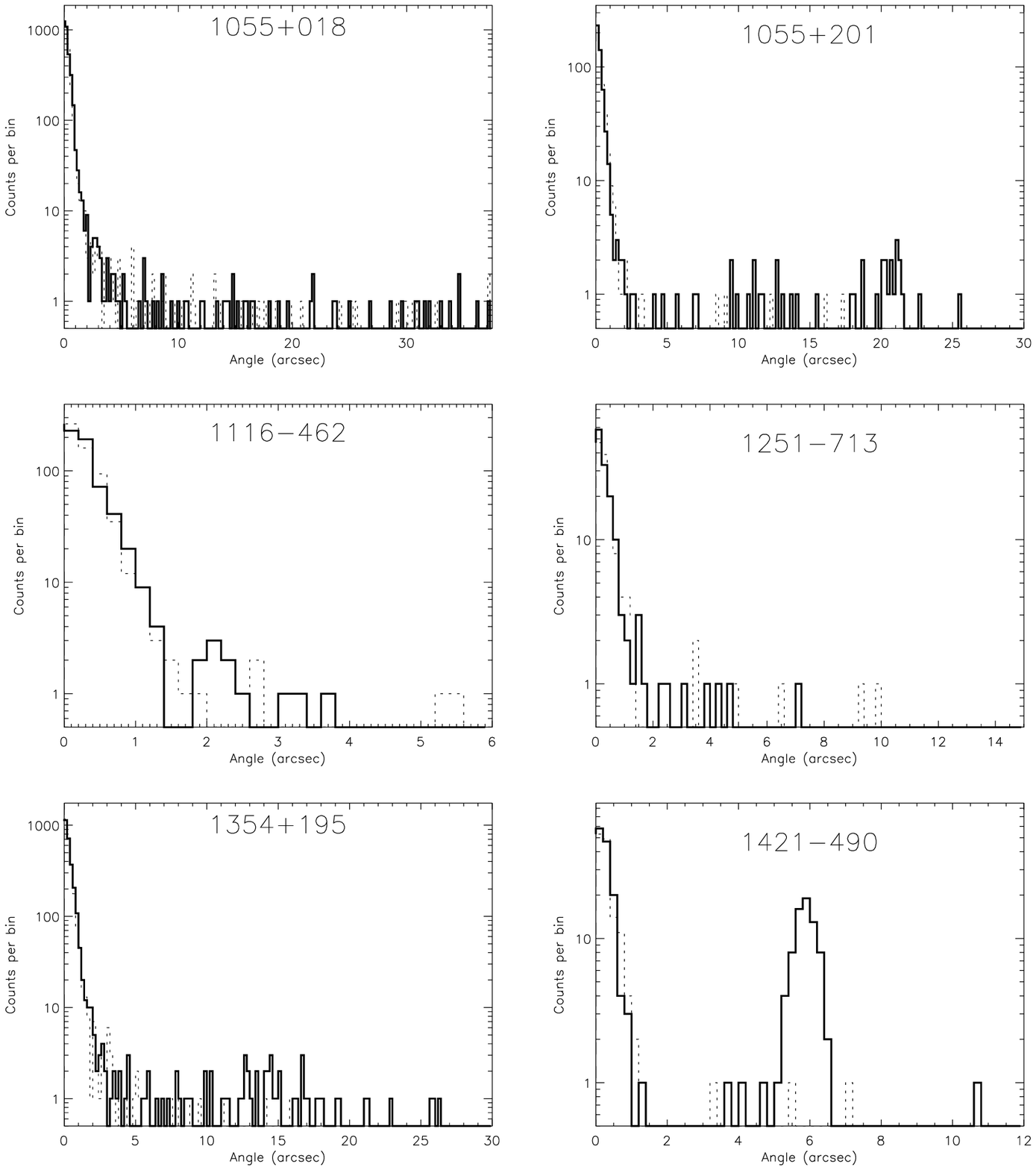}
  \caption{continued.}
\end{figure}

\addtocounter{figure}{-1}

\begin{figure}
\plotone{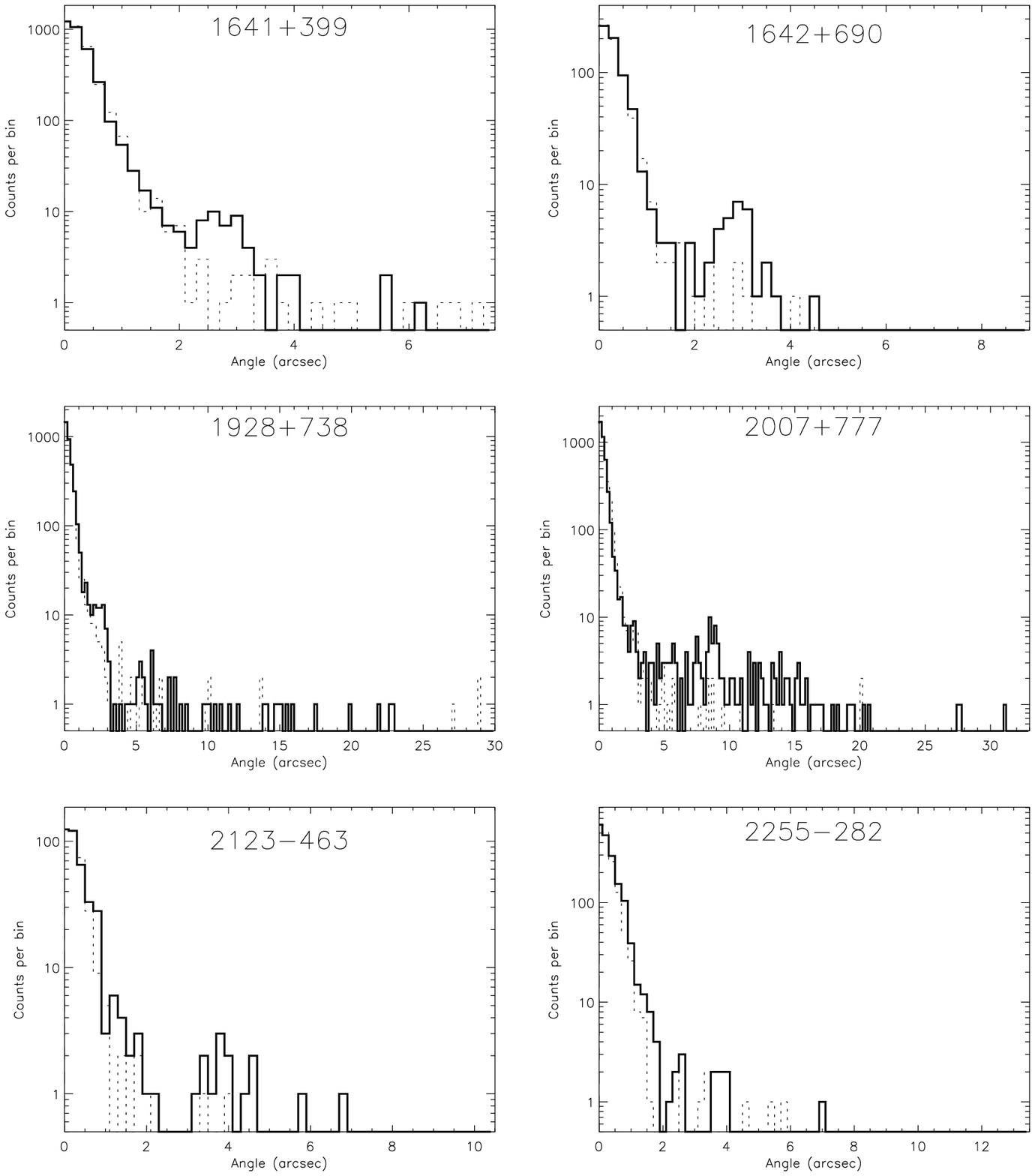}
  \caption{continued.}
\end{figure}

\addtocounter{figure}{-1}

\begin{figure}
\plotone{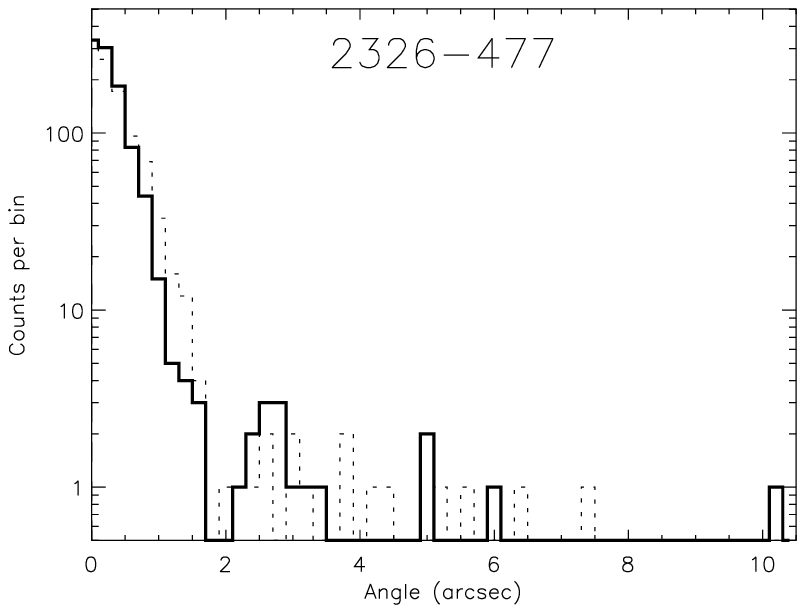}
  \caption{continued.}
\end{figure}

\begin{figure}
\plotone{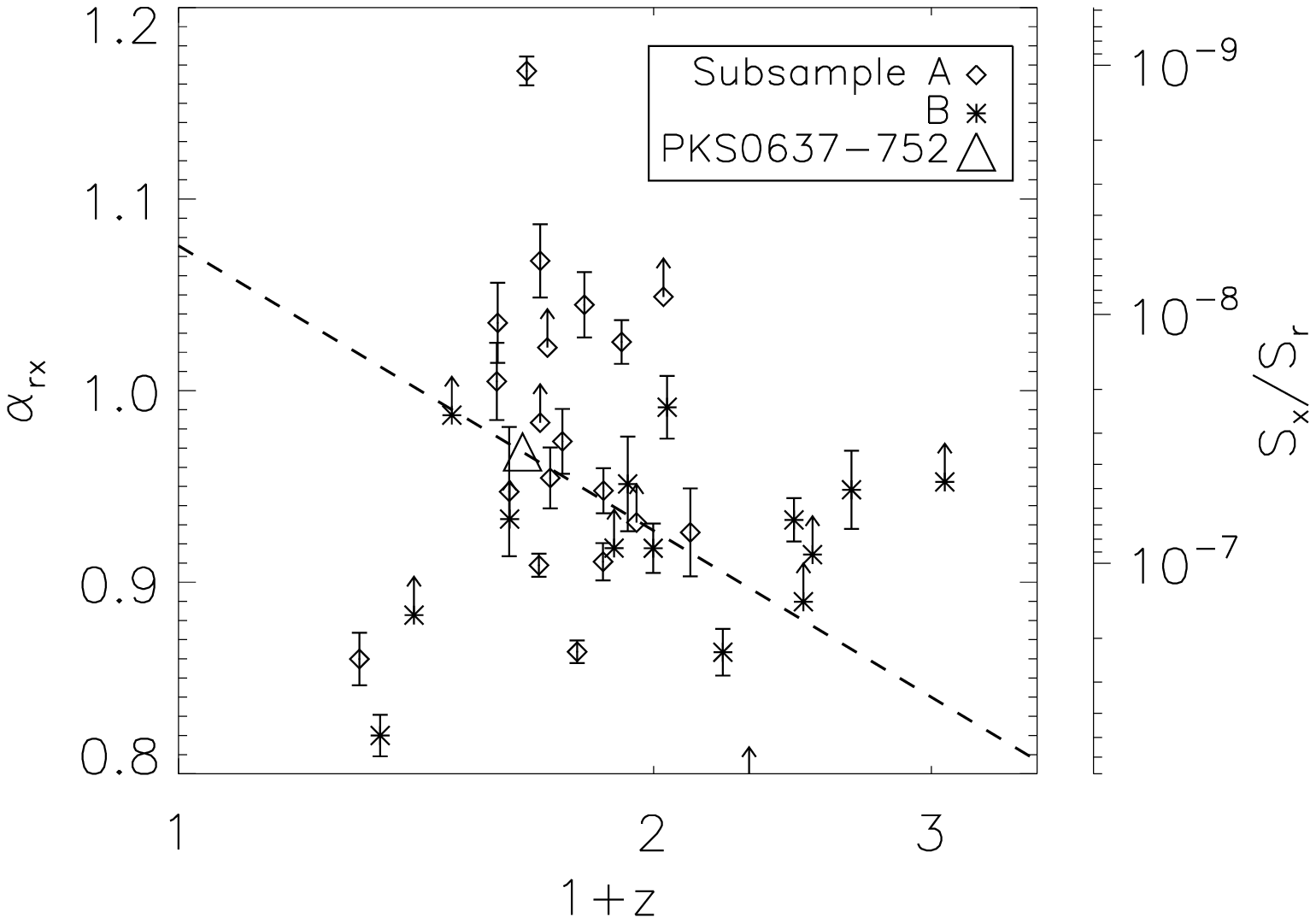}
\caption{Plot of $\alpha_{rx}$ against redshift.
A value of $\alpha_{rx}$ of 1.0 indicates that there is
equal power per logarithmic frequency interval in the X-ray
and radio bands.  The right-hand axis gives the ratio of the
X-ray to radio flux densities, assuming $\nu_r = 8.64 \times 10^9$ Hz
and $\nu_x = 2.4 \times 10^{17}$ Hz.
A change of about 0.13 in $\alpha_{rx}$ results
from a $\times 10$ change in the X-ray flux relative to the radio flux.
The result for PKS 0637-752 is given for comparison.
The dashed line gives the dependence of $\alpha_{rx}$ on $z$ under
the assumptions that the X-ray emission results only from inverse Compton
scattering off of the cosmic microwave background and that the beaming parameters
for all jets are the same as those of PKS~0637$-$752, so that the X-ray to radio
flux density ratio would increase as $(1+z)^{3+\alpha}$ (see Eq.~\ref{eq:r}).
Clearly, there is a wide distribution
of the observed values of $\alpha_{rx}$, indicating that the beaming parameters
vary widely.
\label{fig:alpharx-z} }
\end{figure}

\clearpage

\begin{figure}
\epsscale{0.8}
\plotone{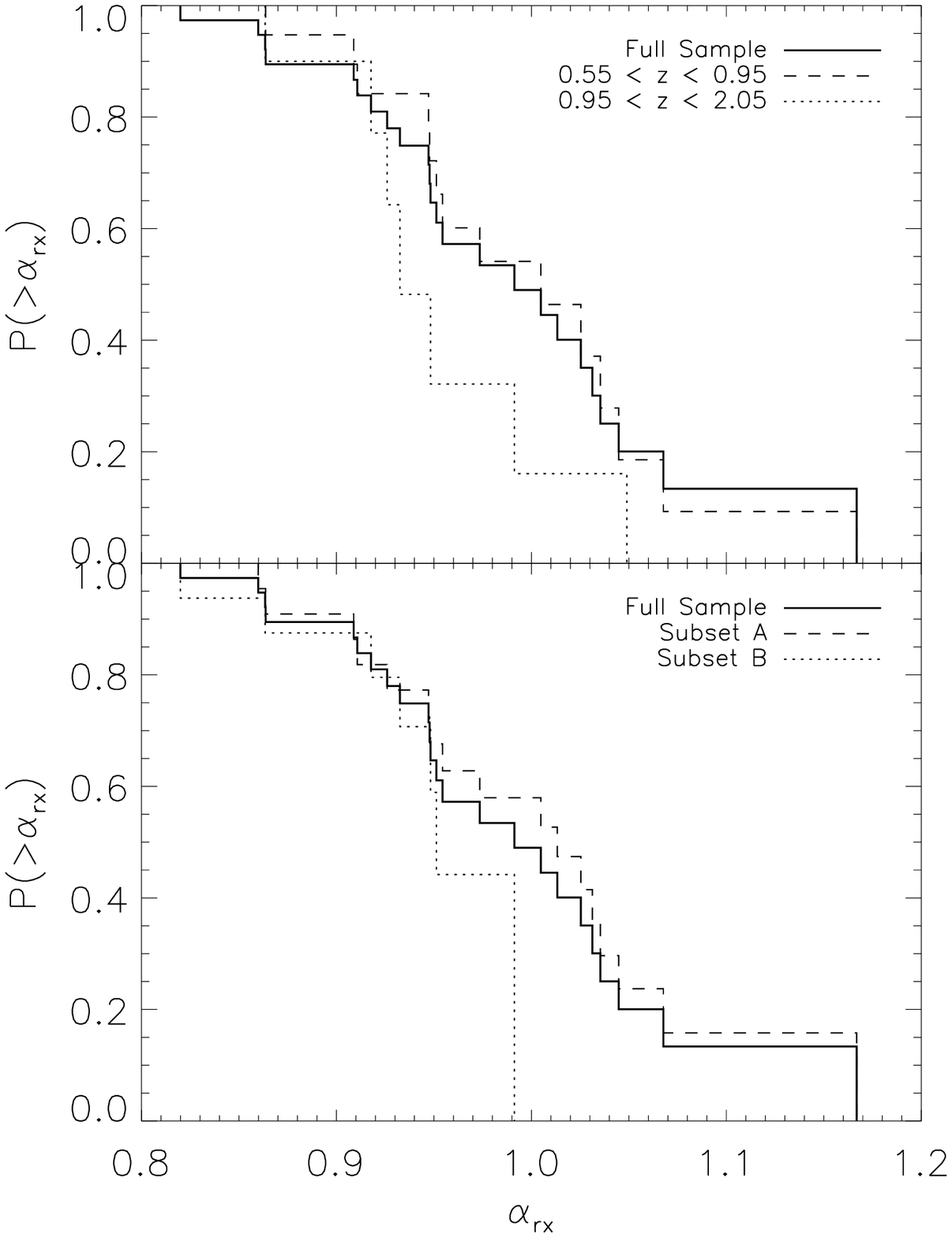}
\caption{Distribution of $\alpha_{rx}$ for 39 sources observed so far in
our sample.  Upper limits are handled by using the Kaplan-Meier
method.  {\it Top:} The sample is divided into two equal groups based on
redshift (excluding those with unknown redshifts or $z < 0.1$).
The high redshift subsample has marginally smaller values of
$\alpha_{rx}$; i.e., the jets' X-ray flux densities are slightly larger relative to their
radio flux densities.
{\it Bottom:} The sample is divided according to the A or B selection
criterion, where A represents a flux-limited subsample and B represents
morphological selection only.  The B subset shows slightly smaller values
of $\alpha_{rx}$ but the sample size is rather small.
\label{fig:arxdistribution} }
\end{figure}

\begin{figure}
\epsscale{1.0}
\includegraphics[angle=90,scale=0.7]{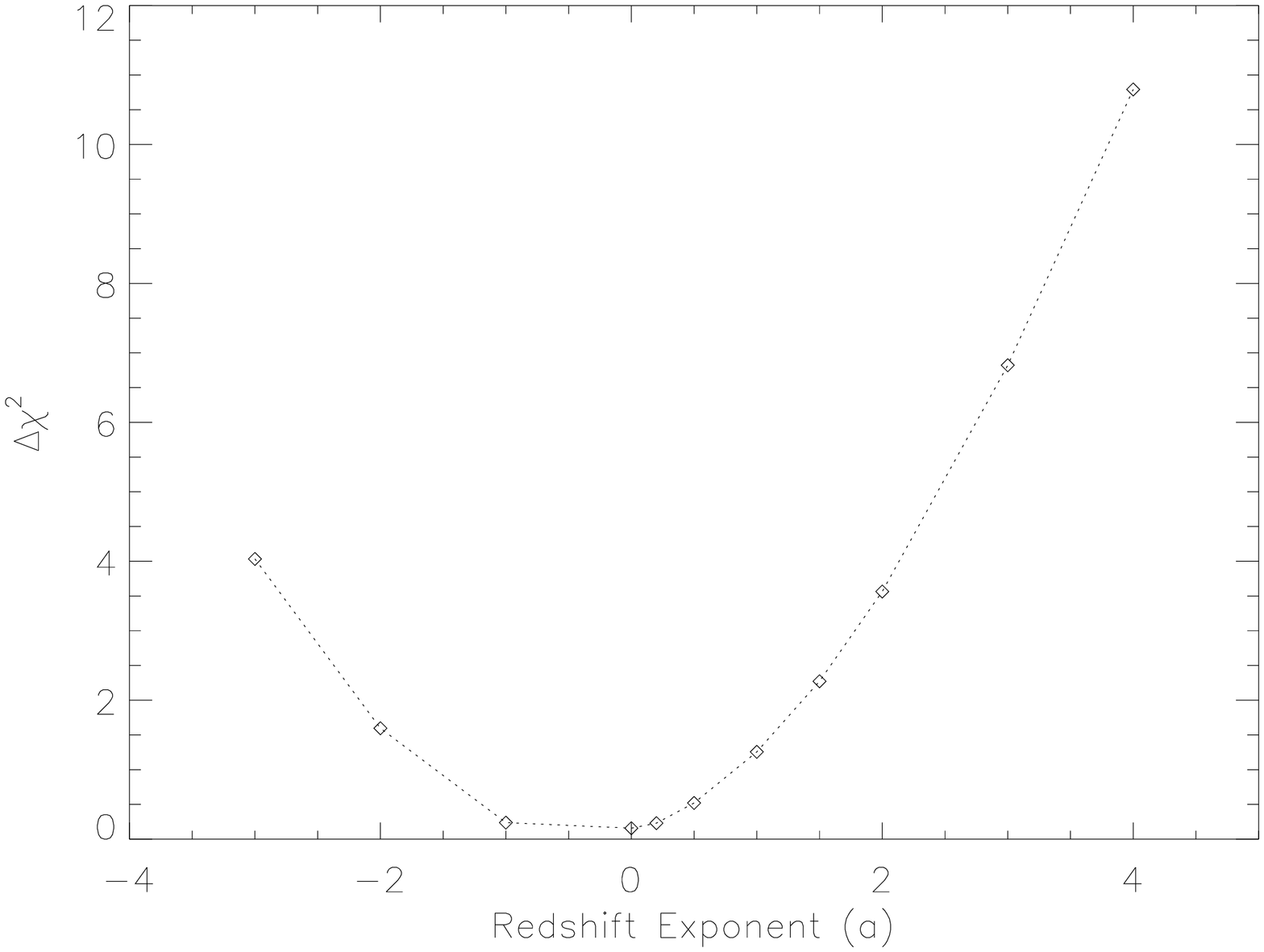}
\caption{Log likelihood dependence on $a$, where $S_x/S_r \propto (1+z)^a$
under the assumption that the distribution of intrinsic magnetic
fields does not depend on redshift.
In the IC-CMB model, $a = 3+\alpha$;
this dependence is ruled out at better
than 99.5\% confidence for $\alpha > 0.5$.  Thus, if the IC-CMB mechanism
is responsible for most of the X-ray emission from quasar jets, then other jet
parameters such as the magnetic field or Lorentz factor must depend on $z$
in a compensatory fashion.
\label{fig:arxtest} }
\end{figure}

\begin{figure}
\includegraphics[angle=90,scale=0.7]{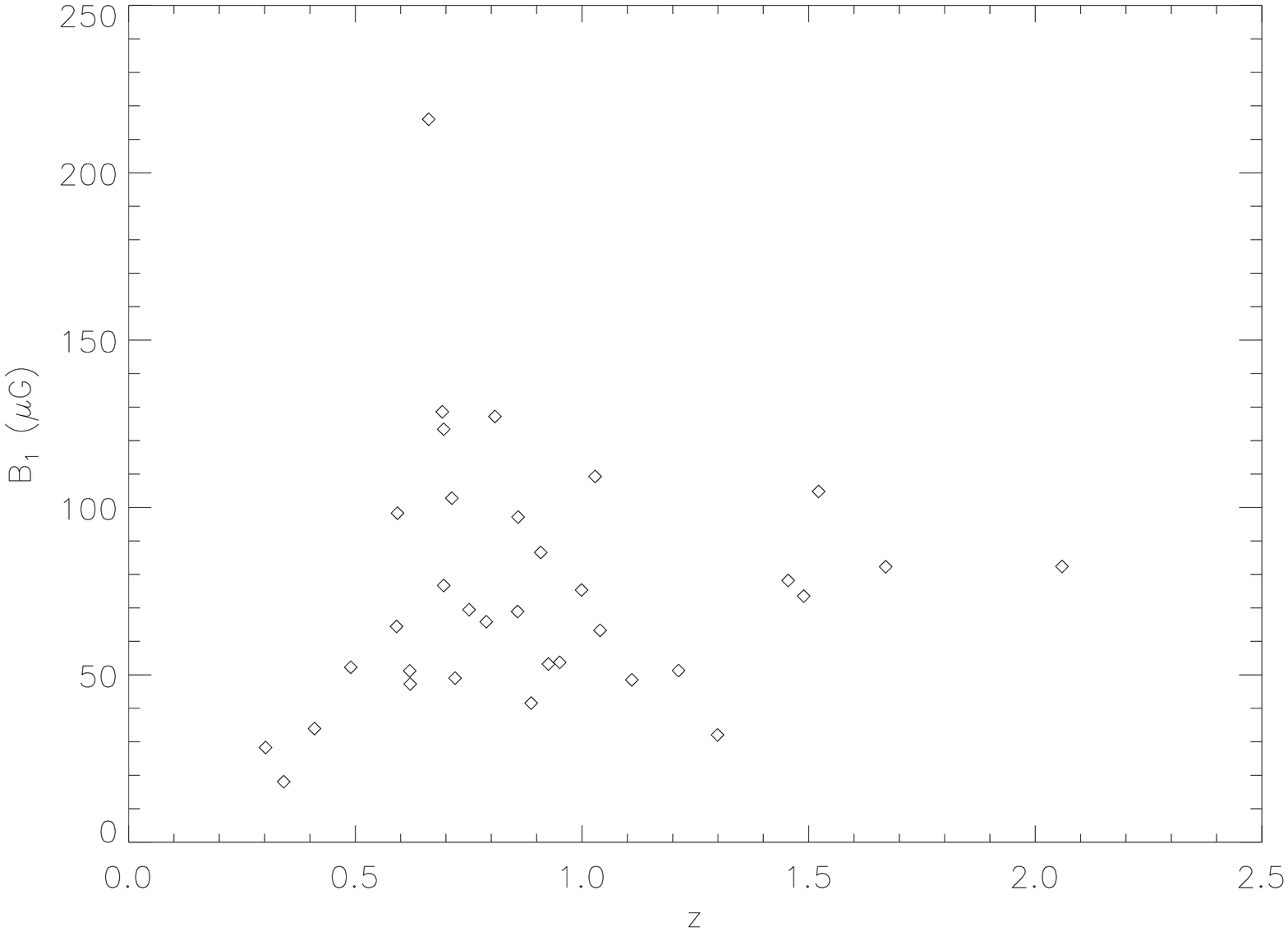}
\caption{Estimated minimum-energy
 values of jet magnetic fields in the absence of
 beaming, $B_1$, plotted against the redshift of the quasar.
 Although the calculation of $B_1$ depends on $z$, there is no
 apparent correlation, probably because of the wide scatter in
 other jet measurements (particularly radio flux and the jet's
 angular length) that go into the calculation of $B_1$. 
\label{fig:b1vz} }
\end{figure}

\begin{figure}
\plotone{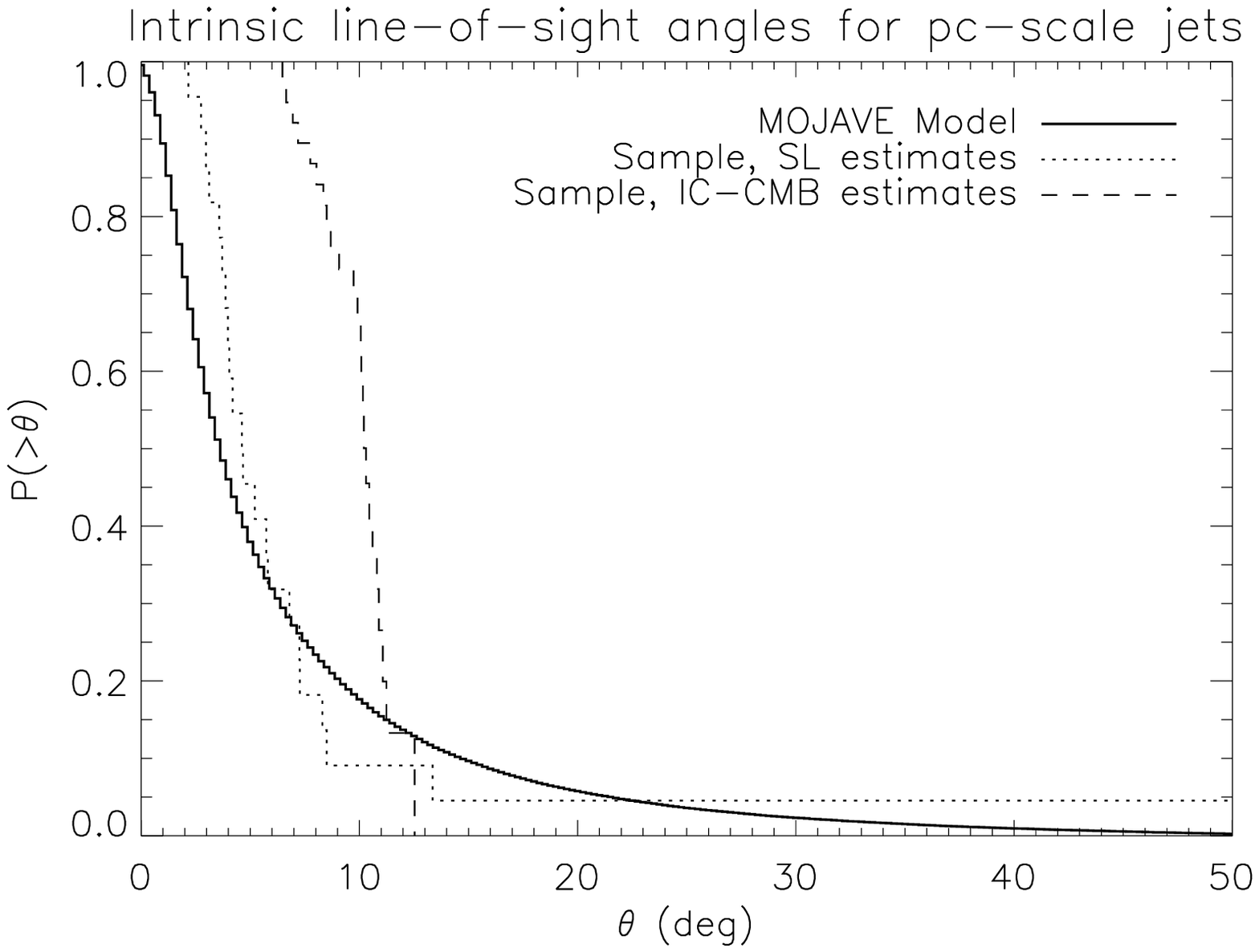}
\caption{Cumulative distributions of angles to the line of sight, $\theta$,
for the model of the
MOJAVE sample \citep{2007ApJ...658..232C} compared to the distribution
of $\theta = 0.5/\beta_{\rm app}$ for our sample based on superluminal (SL)
motion on pc scales (dotted line, where $\beta_{\rm app}$ is
taken from Table~\ref{tab:bends}).
The IC-CMB model is used to derive an angle to the
line of sight as given by eq.~\ref{eq:r}.
\label{fig:sldistribution} }
\end{figure}

\begin{figure}
\plotone{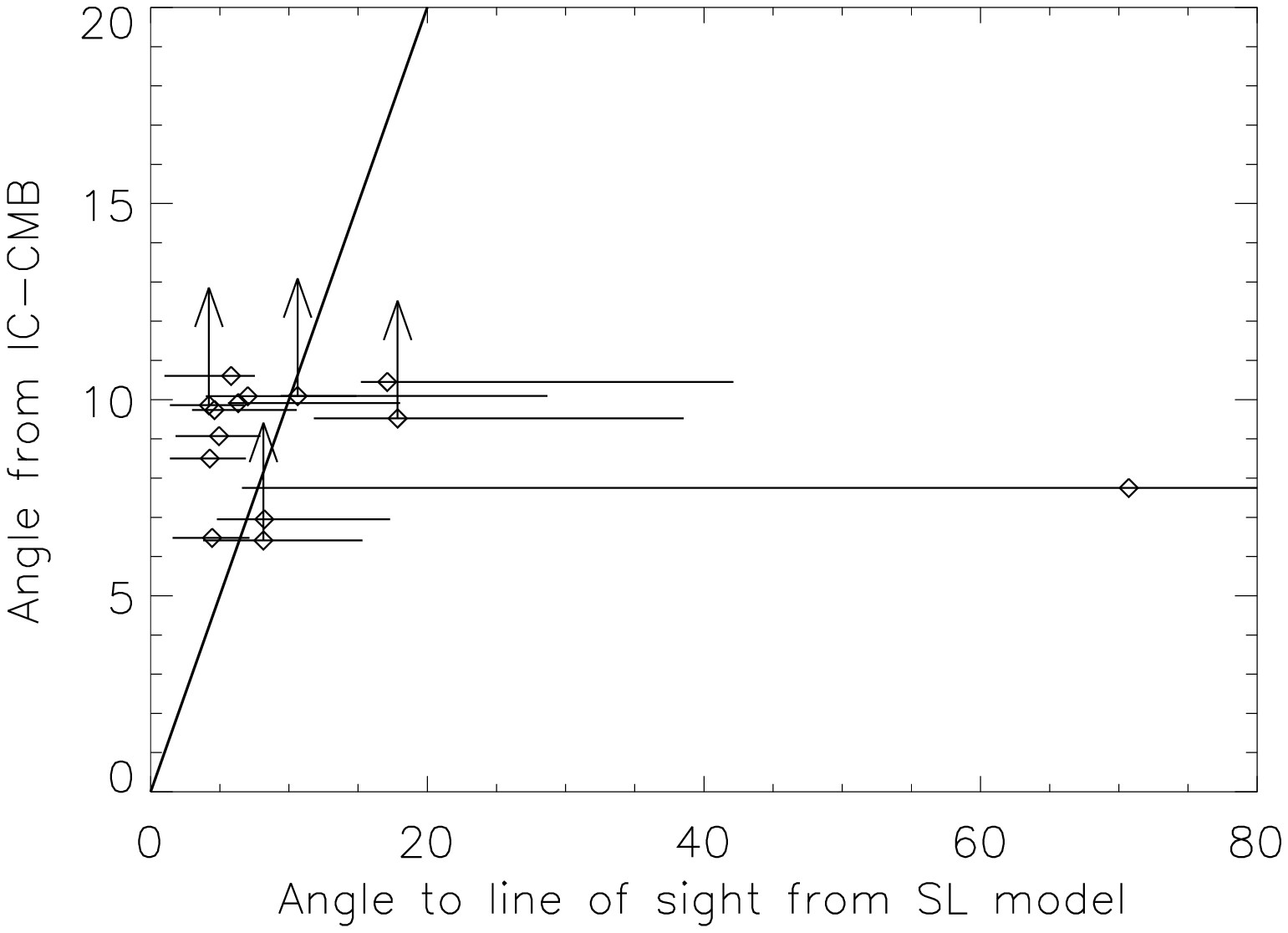}
\caption{Comparison of angles of kpc-scale jets to the line of sight for two computation
methods.  The abscissa is determined from geometric constraints using the apparent
speed of the superluminal (SL) components in the pc-scale jet, combined with the
difference between the position angles of the
pc-scale and kpc-scale jets by the method described in
Appendix~\ref{app:angle}.  The IC-CMB model is used to derive an angle to the
line of sight as given by eq.~\ref{eq:r} to provide the ordinate.  The solid line indicates
where these two angles are equal.  These independent estimates are generally
consistent.  However, there are some notable exceptions, particularly where the angles from
the IC-CMB calculation are $\times$2 larger than those based on geometry and
superluminal motion of the pc-scale jet.  For these exceptions, one may infer that the
jets decelerate substantially from pc scales to kpc scales.
\label{fig:angles} }
\end{figure}

\clearpage

\begin{deluxetable}{lcccc}
\tablecolumns{5}
\tablewidth{0pc}
\tablecaption{{\em Chandra} Observation Log \label{tab:observations} }
\tablehead{
\colhead{Target} & \colhead{{\em Chandra}} & \colhead{Live Time} 
	& \colhead{Date} & \colhead{Ref.\tablenotemark{a}} \\
\colhead{} & \colhead{Obs ID} & \colhead{(s)} 
	& \colhead{(UT)} & }
\startdata
    0234$+$285 &  4898 &  9032 & 2004-06-24 & 1\\
    0454$-$463 &  4893 &  5775 & 2004-06-04 & 1\\
    0820$+$225 &  4897 &  5617 & 2003-12-28 & 1\\
    0923$+$392 &  3048 & 18638 & 2002-10-19 & 1\\
    0954$+$556 &  4842 & 34404 & 2004-06-16 & 2\\
    1040$+$123 &  2136 & 10401 & 2001-02-12 & 3\\
    1055$+$018 &  2137 &  9314 & 2001-01-09 & 4\\
    1055$+$201 &  4889 &  4693 & 2004-01-19 & 1\\
    1116$-$462 &  4891 &  5623 & 2004-03-16 & 1\\
    1251$-$713 &  4892 &  5974 & 2004-03-07 & 1\\
    1354$+$195 &  2140 &  9055 & 2001-01-08 & 4\\
    1421$-$490 &  4895 &  5472 & 2004-01-16 & 1\\
    1641$+$399 &  2143 &  9055 & 2001-04-27 & 3\\
    1642$+$690 &  2142 &  8326 & 2001-03-08 & 3\\
    1928$+$738 &  2145 &  8392 & 2001-04-27 & 3\\
    2007$+$777 &  5709 &  36046 & 2005-05-23 & 5\\
    2123$-$463 &  4890 &  6473 & 2004-03-25 & 1\\
    2255$-$282 &  4894 &  7127 & 2003-11-19 & 1\\
    2326$-$477 &  4896 &  8298 & 2004-06-21 & 1\\
\enddata
\tablenotetext{a}{References refer to previous X-ray imaging
 results: 1) this paper, 2) \cite{2007ApJ...662..900T}, 3) \cite{gambilletal03}, 4) \cite{sambruna04},
 5) \cite{2008ApJ...684..862S}.}
\end{deluxetable}

\begin{deluxetable}{lccrc}
\tablecolumns{5}
\tablewidth{0pc}
\tablecaption{Radio Observations \label{tab:radioCont} }
\tablehead{
\colhead{Target} & \colhead{Instrument} & \colhead{Date}
	& \colhead{Freq.} & \colhead{$5 \times$ RMS noise} \\
\colhead{} & \colhead{} & \colhead{(UT)} 
	& \colhead{(GHz)} & \colhead{(mJy/beam)}}
\startdata
  0234$+$285 &  \mbox{\em VLA} & 2000-11-05 &   1.42 &  3.74 \\
  0454$-$463 & \mbox{\em ATCA} & 2000-05-20 &   8.64 &  2.41 \\
  0820$+$225 &  \mbox{\em VLA} & 2000-11-05 &   1.42 &  5.99 \\
  0923$+$392 &  \mbox{\em VLA} & 2000-11-05 &   4.86 &  6.63 \\
  0954$+$556 &  \mbox{\em VLA} & 2000-11-05 &   4.86 &  4.07 \\
  1040$+$123 &  \mbox{\em VLA} & 1983-09-25 &   4.86 &  1.52 \\
  1055$+$018 &  \mbox{\em VLA} & 1985-05-14 &   4.86 &  1.34 \\
  1055$+$201 &  \mbox{\em VLA} & 1984-12-23 &   1.46 &  3.22 \\
  1116$-$462 & \mbox{\em ATCA} & 2002-02-04 &   8.64 &  1.48 \\
  1251$-$713 & \mbox{\em ATCA} & 1993-07-13 &   4.80 &  6.13 \\
  1354$+$195 &  \mbox{\em VLA} & 1985-04-20 &   4.86 &  1.63 \\
  1421$-$490 & \mbox{\em ATCA} & 2004-05-09 &  17.73 &  1.12 \\
  1641$+$399 &  \mbox{\em VLA} & 1985-01-31 &   4.86 &  3.78 \\
  1642$+$690 &  \mbox{\em VLA} & 1986-05-06 &   4.86 &  1.92 \\
  1928$+$738 &  \mbox{\em VLA} & 1996-11-23 &   1.42 &  1.98 \\
  2007$+$777 &  \mbox{\em VLA} & 2000-11-05 &   1.42 &  1.70 \\
  2123$-$463 & \mbox{\em ATCA} & 2004-05-10 &  17.73 &  1.08 \\
  2255$-$282 &  \mbox{\em VLA} & 2000-11-05 &   4.86 &  2.05 \\
  2326$-$477 & \mbox{\em ATCA} & 2002-01-31 &   8.64 &  1.43 \\
\enddata
\end{deluxetable}

\begin{deluxetable}{rrrrrrrrrrr}
\tablecolumns{11}
\tablewidth{0pc}
\tabletypesize{\scriptsize}
\tablecaption{Quasar Jet Measurements \label{tab:jetresults} }
\tablehead{
\colhead{Target} & \colhead{PA} 
	& \colhead{$\theta_i$} & \colhead{$\theta_o$} 
	& \colhead{$S_r$\tablenotemark{a}} & \colhead{$\nu_r$} & \colhead{Count Rate} & \colhead{$S_x$\tablenotemark{a}} 
	& \colhead{$\alpha_{rx}$} & \colhead{$P_{jet}$\tablenotemark{b}} & \colhead{X?\tablenotemark{c}} \\
\colhead{ } & \colhead{(\arcdeg)} 
	& \colhead{(\arcsec)} & \colhead{(\arcsec)} 
	& \colhead{(mJy)} & \colhead{(GHz)} & \colhead{(10$^{-3}$ cps)} & \colhead{(nJy)} & \colhead{} 
	& \colhead{} & \colhead{} }
\startdata
0234$+$285 &  -20 &  1.5 & 10.0 &   66.0 $\pm$ 4.5 &  1.42 &   5.20 $\pm$  1.15 &      5.2 &     0.86 $\pm$ 0.01 & $<$ 1e$-$10 & Y \\
0454$-$463 &  150 &  1.5 &  8.0 &   62.4 $\pm$ 2.2 &  8.64 &  10.39 $\pm$  1.70 &     10.4 &     0.91 $\pm$ 0.01 & $<$ 1e$-$10 & Y \\
0820$+$225 &  -90 &  1.5 &  5.0 &   50.4 $\pm$ 3.3 &  1.42 &   0.18 $\pm$  0.31 & $<$  1.1 & $>$ 0.93            & 2.64e-01 & N \\
0923$+$392 &   75 &  1.5 &  4.0 &   74.1 $\pm$ 9.7 &  4.86 &   0.16 $\pm$  0.62 & $<$  2.0 & $>$ 0.98            & 3.55e-01 & N \\
0954$+$556 &  -60 &  1.5 &  5.0 &  108.0 $\pm$ 7.1 &  4.86 &   1.40 $\pm$  0.27 &      1.4 &     1.03 $\pm$ 0.01 & $<$ 1e$-$10 & Y \\
1040$+$123 &  -90 &  1.5 &  6.0 &  253.8 $\pm$ 4.2 &  4.86 &   0.00 $\pm$  0.72 & $<$  2.2 & $>$ 1.05            & 5.25e-01 & N \\
1055$+$018 &  180 &  1.5 & 25.0 &   57.7 $\pm$ 2.6 &  4.86 &   0.54 $\pm$  1.49 & $<$  5.0 & $>$ 0.92            & 3.03e-01 & N \\
1055$+$201 &  -10 &  1.5 & 20.0 &  137.8 $\pm$ 4.1 &  1.46 &   3.41 $\pm$  1.48 &      3.4 &     0.93 $\pm$ 0.02 & 2.76e-04 & Y \\
1116$-$462 &  -85 &  1.5 &  4.0 &  126.2 $\pm$ 0.6 &  8.64 &   0.89 $\pm$  0.73 & $<$  3.1 & $>$ 1.02            & 4.26e-02 & N \\
1251$-$713 &  180 &  1.5 & 10.0 &   35.7 $\pm$ 2.6 &  4.80 &   0.33 $\pm$  0.63 & $<$  2.2 & $>$ 0.94            & 2.56e-01 & N \\
1354$+$195 &  165 &  1.5 & 20.0 &  109.4 $\pm$ 2.3 &  4.86 &   4.97 $\pm$  1.39 &      5.0 &     0.95 $\pm$ 0.02 & $<$ 1e$-$10 & Y \\
1421$-$490 &   30 &  1.5 &  8.0 & 2720.0 $\pm$ 1.8 & 17.73 &  12.97 $\pm$  1.60 &     13.0 &     1.17 $\pm$ 0.01 & $<$ 1e$-$10 & Y \\
1641$+$399 &  -30 &  1.5 &  5.0 &  295.9 $\pm$ 5.1 &  4.86 &   3.20 $\pm$  1.18 &      3.2 &     1.04 $\pm$ 0.02 & 4.89e-06 & Y \\
1642$+$690 &  170 &  1.5 &  6.0 &   85.3 $\pm$ 2.9 &  4.86 &   2.76 $\pm$  0.82 &      2.8 &     0.97 $\pm$ 0.02 & 5.23e-08 & Y \\
1928$+$738 &  170 &  1.5 & 20.0 &   80.5 $\pm$ 3.0 &  1.42 &   6.79 $\pm$  1.75 &      6.8 &     0.86 $\pm$ 0.01 & $<$ 1e$-$10 & Y \\
2007$+$777 & -105 &  1.5 & 22.0 &   17.9 $\pm$ 2.4 &  1.42 &   3.22 $\pm$  0.51 &      3.2 &     0.82 $\pm$ 0.01 & $<$ 1e$-$10 & Y \\
2123$-$463 &  100 &  1.5 &  7.0 &   14.3 $\pm$ 1.5 & 17.73 &   2.47 $\pm$  0.79 &      2.5 &     0.95 $\pm$ 0.02 & 8.11e-08 & Y \\
2255$-$282 &  -70 &  1.5 &  9.0 &   40.9 $\pm$ 3.6 &  4.86 &   1.96 $\pm$  0.84 &      2.0 &     0.95 $\pm$ 0.02 & 1.99e-04 & Y \\
2326$-$477 & -105 &  1.5 &  7.0 &   12.7 $\pm$ 0.8 &  8.64 &  -0.12 $\pm$  0.73 & $<$  2.1 & $>$ 0.79            & 6.22e-01 & N \\
\enddata
\tablenotetext{a}{The jet radio flux density is measured at $\nu_r$ for the same region as for
the X-ray count rate, given by the PA, $\theta_i$, and $\theta_o$ parameters.  The X-ray flux
density is given at 1 keV assuming a conversion of 1 $\mu$Jy/(count/s), which is good
to $\sim$ 10\% for power law spectra with low column densities and X-ray spectral indices
near 0.5.}
\tablenotetext{b}{The quantity $P_{jet}$ is defined as the chance that there are more
counts than observed in the specified region under the null hypothesis that the counts
are background events.}
\tablenotetext{c}{The jet is defined to be detected if $P_{jet} < 0.0026$ (see text).}
\end{deluxetable}

\begin{deluxetable}{rrrrrrrrrrr}
\tablecolumns{11}
\tablewidth{0pc}
\tabletypesize{\scriptsize}
\tablecaption{Quasar Knot Measurements\tablenotemark{a} \label{tab:knotresults} }
\tablehead{
\colhead{Target} & \colhead{PA} 
	& \colhead{$\theta_i$} & \colhead{$\theta_o$} 
	& \colhead{$S_r$} & \colhead{$\nu_r$} & \colhead{Count Rate} & \colhead{$S_x$} 
	& \colhead{$\alpha_{rx}$} & \colhead{$P_{knot}$} & \colhead{X?} \\
\colhead{ } & \colhead{(\arcdeg)} 
	& \colhead{(\arcsec)} & \colhead{(\arcsec)} 
	& \colhead{(mJy)} & \colhead{(GHz)} & \colhead{(10$^{-3}$ cps)} & \colhead{(nJy)} & \colhead{} 
	& \colhead{} & \colhead{} }
\startdata
0234$+$285 &    0 &  1.5 &  3.5 &   35.6 $\pm$ 1.4 &  1.42 &   4.32 $\pm$  0.88 &      4.3 &     0.84 $\pm$ 0.01 & $<$ 1e$-$10 & Y \\
0234$+$285 &    0 &  4.5 &  6.5 &    3.1 $\pm$ 1.4 &  1.42 &   0.44 $\pm$  0.27 &      0.4 &     0.83 $\pm$ 0.04 & 3.66e-03 & Y \\
0454$-$463 &  125 &  1.5 &  3.5 &   38.7 $\pm$ 0.8 &  8.64 &  10.91 $\pm$  1.48 &     10.9 &     0.88 $\pm$ 0.01 & $<$ 1e$-$10 & Y \\
0820$+$225 &  -90 &  1.5 &  3.5 &   50.2 $\pm$ 2.0 &  1.42 &   0.00 $\pm$  0.25 & $<$  0.8 & $>$ 0.95            & 6.32e-01 & N \\
0923$+$392 &  -80 &  1.5 &  3.5 &   -1.0 $\pm$ 7.1 &  4.86 &  -0.43 $\pm$  0.58 & $<$  1.3 & $>$ 0.91            & 8.45e-01 & N \\
0954$+$556 &  -60 &  1.5 &  3.5 &   95.8 $\pm$ 4.4 &  4.86 &   1.34 $\pm$  0.23 &      1.3 &     1.02 $\pm$ 0.01 & $<$ 1e$-$10 & Y \\
1040$+$123 &  -80 &  1.5 &  3.5 &   41.2 $\pm$ 2.3 &  4.86 &   0.00 $\pm$  0.53 & $<$  1.6 & $>$ 0.96            & 5.34e-01 & N \\
1055$+$018 &  180 &  1.5 &  3.5 &    6.7 $\pm$ 0.3 &  4.86 &   0.11 $\pm$  0.69 & $<$  2.2 & $>$ 0.84            & 4.40e-01 & N \\
1055$+$201 &    0 &  1.5 &  3.5 &   18.5 $\pm$ 0.8 &  1.46 &   0.21 $\pm$  0.48 & $<$  1.6 & $>$ 0.86            & 3.23e-01 & N \\
1055$+$201 &    0 &  9.0 & 13.0 &   10.5 $\pm$ 1.1 &  1.46 &   1.07 $\pm$  0.56 &      1.1 &     0.85 $\pm$ 0.03 & 5.94e-04 & Y \\
1116$-$462 &  -85 &  1.5 &  3.5 &  110.8 $\pm$ 0.4 &  8.64 &   0.53 $\pm$  0.64 & $<$  2.5 & $>$ 1.03            & 1.33e-01 & N \\
1251$-$713 &  180 &  1.5 &  3.5 &    1.7 $\pm$ 0.6 &  4.80 &   0.33 $\pm$  0.33 & $<$  1.3 & $>$ 0.79            & 8.03e-02 & N \\
1354$+$195 &  163 &  2.0 &  4.0 &   24.3 $\pm$ 0.6 &  4.86 &  -0.55 $\pm$  0.65 & $<$  1.4 & $>$ 0.94            & 8.95e-01 & N \\
1354$+$195 &  163 &  4.0 &  6.0 &   13.4 $\pm$ 0.6 &  4.86 &   0.44 $\pm$  0.35 & $<$  1.5 & $>$ 0.90            & 3.35e-02 & N \\
1354$+$195 &  163 &  6.0 &  8.0 &   12.0 $\pm$ 0.6 &  4.86 &   0.33 $\pm$  0.29 & $<$  1.2 & $>$ 0.91            & 5.26e-02 & N \\
1354$+$195 &  163 &  8.0 & 10.0 &   13.0 $\pm$ 0.6 &  4.86 &   0.22 $\pm$  0.31 & $<$  1.2 & $>$ 0.92            & 1.85e-01 & N \\
1354$+$195 &  163 & 10.0 & 12.0 &    5.4 $\pm$ 0.6 &  4.86 &   0.44 $\pm$  0.27 &      0.4 &     0.92 $\pm$ 0.04 & 3.66e-03 & Y \\
1354$+$195 &  163 & 12.0 & 14.0 &    7.9 $\pm$ 0.6 &  4.86 &   0.99 $\pm$  0.37 &      1.0 &     0.90 $\pm$ 0.02 & 1.11e-07 & Y \\
1354$+$195 &  163 & 14.0 & 16.0 &    7.3 $\pm$ 0.6 &  4.86 &   0.77 $\pm$  0.37 &      0.8 &     0.91 $\pm$ 0.03 & 2.37e-04 & Y \\
1421$-$490 &   30 &  5.0 &  7.0 & 2704.6 $\pm$ 0.8 & 17.73 &  12.24 $\pm$  1.52 &     12.2 &     1.17 $\pm$ 0.01 & $<$ 1e$-$10 & Y \\
1641$+$399 &  -35 &  1.5 &  3.5 &  263.0 $\pm$ 3.2 &  4.86 &   3.64 $\pm$  0.99 &      3.6 &     1.02 $\pm$ 0.02 & $<$ 1e$-$10 & Y \\
1642$+$690 &  175 &  2.0 &  4.0 &   59.9 $\pm$ 1.6 &  4.86 &   2.88 $\pm$  0.66 &      2.9 &     0.95 $\pm$ 0.01 & $<$ 1e$-$10 & Y \\
1928$+$738 & -170 &  2.0 &  4.0 &   16.1 $\pm$ 0.5 &  1.42 &   4.65 $\pm$  0.98 &      4.6 &     0.79 $\pm$ 0.01 & $<$ 1e$-$10 & Y \\
2007$+$777 & -105 &  4.0 &  6.0 &    0.9 $\pm$ 0.5 &  1.42 &   0.50 $\pm$  0.15 &      0.5 &     0.77 $\pm$ 0.03 & 2.48e-08 & Y \\
2007$+$777 & -105 &  7.5 &  9.5 &    3.3 $\pm$ 0.5 &  1.42 &   0.97 $\pm$  0.17 &      1.0 &     0.79 $\pm$ 0.01 & $<$ 1e$-$10 & Y \\
2007$+$777 & -105 & 11.0 & 13.0 &    1.2 $\pm$ 0.5 &  1.42 &   0.25 $\pm$  0.09 &      0.2 &     0.81 $\pm$ 0.03 & 1.11e-07 & Y \\
2007$+$777 & -110 & 13.0 & 15.0 &    0.7 $\pm$ 0.5 &  1.42 &   0.22 $\pm$  0.09 &      0.2 &     0.81 $\pm$ 0.03 & 1.13e-06 & Y \\
2007$+$777 & -105 & 15.0 & 17.0 &    1.3 $\pm$ 0.5 &  1.42 &   0.28 $\pm$  0.10 &      0.3 &     0.81 $\pm$ 0.03 & 1.00e-08 & Y \\
2123$-$463 &  112 &  1.5 &  3.5 &    2.6 $\pm$ 0.7 & 17.73 &   1.24 $\pm$  0.58 &      1.2 &     0.89 $\pm$ 0.03 & 2.92e-04 & Y \\
2123$-$463 &  112 &  3.0 &  5.0 &    8.0 $\pm$ 0.7 & 17.73 &   1.39 $\pm$  0.51 &      1.4 &     0.95 $\pm$ 0.02 & 1.11e-07 & Y \\
2255$-$282 &  -70 &  1.5 &  3.5 &   14.5 $\pm$ 1.5 &  4.86 &   1.26 $\pm$  0.70 &      1.3 &     0.92 $\pm$ 0.03 & 3.72e-03 & Y \\
2326$-$477 & -110 &  1.5 &  3.5 &   -7.0 $\pm$ 0.4 &  8.64 &   0.48 $\pm$  0.61 & $<$  2.3 & $>$ 0.75            & 1.46e-01 & N \\
\enddata
\tablenotetext{a}{All quantities are defined as in Table~\ref{tab:jetresults}, except that knots are defined
to be detected if $P_{knot} < 0.0062$, which gives a 20\% chance of one false detection in 32 trials.}
\end{deluxetable}

\begin{deluxetable}{rcrrrrrrr}
\tablecolumns{7}
\tablewidth{0pc}
\tabletypesize{\scriptsize}
\tablecaption{Jet Beaming Model Parameters \label{tab:beaming} }
\tablehead{
\colhead{Target} & \colhead{z} &  \colhead{A/B} & \colhead{$\alpha_{rx}$} & \colhead{$R_1$\tablenotemark{a}}
	& \colhead{$V$\tablenotemark{b}} & \colhead{$B_1$\tablenotemark{c}} &
	\colhead{$K$\tablenotemark{d}} & \colhead{$\theta$\tablenotemark{e}} \\
\colhead{ } & \colhead{}  & \colhead{}  & \colhead{}  & \colhead{($10^{-3}$)} & \colhead{(pc$^3$)} &
	\colhead{($\mu$G)} & \colhead{} & \colhead{(\arcdeg)} }
\startdata
0208$-$512 &  0.999 & B &     0.92 &       132.8 &  1.0e+12 &   75. &      23.6 &        9 \\
0229$+$131 &  2.059 & B & $>$ 0.95 &  $<$   55.8 &  1.2e+12 &   82. & $<$   6.5 &  $>$  13 \\
0234$+$285 &  1.213 & B &     0.86 &       300.5 &  2.2e+12 &   51. &      20.4 &        9 \\
0413$-$210 &  0.808 & A &     1.04 &        13.0 &  5.7e+11 &  127. &      13.6 &       10 \\
0454$-$463 &  0.858 & A &     0.91 &       149.8 &  1.3e+12 &   69. &      27.0 &        8 \\
0745$+$241 &  0.410 & B & $>$ 0.88 &  $<$  230.3 &  4.7e+11 &   34. & $<$  30.2 &  $>$   8 \\
0820$+$225 &  0.951 & A & $>$ 0.93 &  $<$   83.3 &  7.7e+11 &   54. & $<$  13.7 &  $>$  10 \\
0858$-$771 &  0.490 & B & $>$ 0.99 &  $<$   40.3 &  5.4e+11 &   52. & $<$  15.7 &  $>$  10 \\
0903$-$573 &  0.695 & A &     1.07 &        10.1 &  6.4e+11 &  123. &      13.1 &       10 \\
0920$-$397 &  0.591 & A &     1.00 &        29.8 &  1.4e+12 &   64. &      14.3 &       10 \\
0923$+$392 &  0.695 & A & $>$ 0.98 &  $<$   38.8 &  4.0e+11 &   77. & $<$  17.2 &  $>$  10 \\
0954$+$556 &  0.909 & A &     1.03 &        18.4 &  7.4e+11 &   87. &      10.0 &       11 \\
1030$-$357 &  1.455 & B &     0.93 &       103.0 &  3.9e+12 &   78. &      13.8 &       10 \\
1040$+$123 &  1.029 & A & $>$ 1.05 &  $<$   12.1 &  1.0e+12 &  109. & $<$   8.8 &  $>$  12 \\
1046$-$409 &  0.620 & A &     0.95 &        80.0 &  6.2e+11 &   51. &      18.9 &        9 \\
1055$+$018 &  0.888 & B & $>$ 0.92 &  $<$  123.9 &  4.9e+12 &   42. & $<$  14.2 &  $>$  10 \\
1055$+$201 &  1.110 & A &     0.93 &        92.2 &  4.5e+12 &   49. &      11.1 &       11 \\
1116$-$462 &  0.713 & A & $>$ 1.02 &  $<$   22.0 &  4.1e+11 &  103. & $<$  16.5 &  $>$  10 \\
1202$-$262 &  0.789 & A &     0.86 &       335.5 &  1.2e+12 &   66. &      43.7 &        7 \\
1354$+$195 &  0.720 & A &     0.95 &        64.8 &  3.1e+12 &   49. &      14.2 &       10 \\
1421$-$490 &  0.662 & A &     1.17 &         2.4 &  9.8e+11 &  216. &      10.8 &       11 \\
1424$-$418 &  1.522 & B & $>$ 0.91 &  $<$  140.6 &  8.1e+11 &  105. & $<$  20.8 &  $>$   9 \\
1641$+$399 &  0.593 & A &     1.04 &        15.4 &  4.6e+11 &   98. &      15.1 &       10 \\
1642$+$690 &  0.751 & A &     0.97 &        46.2 &  7.9e+11 &   69. &      16.0 &       10 \\
1655$+$077 &  0.621 & B & $>$ 0.93 &  $<$   94.7 &  4.9e+11 &   47. & $<$  19.1 &  $>$   9 \\
1828$+$487 &  0.692 & A &     0.91 &       145.3 &  2.4e+11 &  129. &      60.3 &        6 \\
1928$+$738 &  0.302 & A &     0.86 &       321.3 &  7.4e+11 &   28. &      35.9 &        8 \\
2007$+$777 &  0.342 & B &     0.82 &       685.0 &  1.0e+12 &   18. &      32.7 &        8 \\
2052$-$474 &  1.489 & B & $>$ 0.89 &  $<$  214.4 &  6.7e+11 &   74. & $<$  18.9 &  $>$   9 \\
2101$-$490 &  1.040 & B &     0.99 &        37.6 &  3.4e+12 &   63. &       9.4 &       11 \\
2123$-$463 &  1.670 & B &     0.95 &        87.6 &  1.5e+12 &   82. &      11.1 &       11 \\
2251$+$158 &  0.859 & A &     0.95 &        72.9 &  1.1e+12 &   97. &      25.5 &        8 \\
2255$-$282 &  0.926 & B &     0.95 &        68.5 &  1.6e+12 &   53. &      12.5 &       10 \\
2326$-$477 &  1.299 & B & $>$ 0.79 &  $<$ 1111.1 &  1.4e+12 &   32. & $<$  24.4 &  $>$   9 \\
\enddata
\tablenotetext{a}{The ratio of the inverse Compton to synchrotron
   luminosities; see Paper I.}
\tablenotetext{b}{$V$ is the volume of the synchrotron emission region.  Note that the
values reported in Paper I are incorrect and are corrected here.}
\tablenotetext{c}{$B_1$ is the minimum energy magnetic field; see Paper I.}
\tablenotetext{d}{$K$ is a function of observable and assumed quantities;
  large values indicate stronger beaming in the IC-CMB model.  See Paper I
  for details.}
\tablenotetext{e}{The bulk Lorentz factor is assumed to  be 15.}
\end{deluxetable}

\begin{deluxetable}{rrrrrrr}
\tablecolumns{7}
\tablewidth{0pc}
\tabletypesize{\scriptsize}
\tablecaption{Quasar Knot Beaming Model Parameters\tablenotemark{a} \label{tab:knotbeaming} }
\tablehead{
\colhead{Target} & \colhead{$\alpha_{rx}$} & \colhead{$R_1$}
	& \colhead{$V$} & \colhead{$B_1$} &
	\colhead{$K$} & \colhead{$\theta$} \\
\colhead{ } & \colhead{}  & \colhead{($10^{-3}$)} & \colhead{(pc$^3$)} &
	\colhead{($\mu$G)} & \colhead{} & \colhead{(\arcdeg)} }
\startdata
  0234$+$285 &     0.84 &       461.6 &  509. &   65. &      32.9 &        9 \\
  0454$-$463 &     0.88 &       254.0 &  403. &   84. &      44.2 &        8 \\
  0820$+$225 & $>$ 0.95 &  $<$   57.3 &  440. &   63. & $<$  13.0 &  $>$  12 \\
  0923$+$392 & $>$ 0.85 &  $<$  416.8 &  321. &   37. & $<$  30.7 &  $>$   9 \\
  0954$+$556 &     1.00 &        24.8 &  424. &   92. &      12.6 &       12 \\
  1040$+$123 & $>$ 0.96 &  $<$   54.7 &  465. &   82. & $<$  15.2 &  $>$  12 \\
  1055$+$018 & $>$ 0.84 &  $<$  461.5 &  415. &   45. & $<$  32.1 &  $>$   9 \\
  1055$+$201 & $>$ 0.88 &  $<$  212.2 &  487. &   59. & $<$  21.3 &  $>$  10 \\
  1055$+$201 &     0.85 &       363.9 &  975. &   36. &      17.8 &       11 \\
  1116$-$462 & $>$ 0.75 &  $<$ 2466.4 &  331. &   27. & $<$  58.8 &  $>$   7 \\
  1354$+$195 & $>$ 0.94 &  $<$   82.6 &  335. &   60. & $<$  19.9 &  $>$  11 \\
  1354$+$195 & $>$ 0.90 &  $<$  158.2 &  335. &   51. & $<$  24.1 &  $>$  10 \\
  1354$+$195 & $>$ 0.91 &  $<$  143.3 &  335. &   49. & $<$  22.1 &  $>$  10 \\
  1354$+$195 & $>$ 0.92 &  $<$  127.0 &  335. &   50. & $<$  21.2 &  $>$  10 \\
  1354$+$195 &     0.92 &       117.8 &  335. &   39. &      15.8 &       12 \\
  1354$+$195 &     0.90 &       179.2 &  335. &   44. &      22.2 &       10 \\
  1354$+$195 &     0.91 &       151.0 &  335. &   43. &      19.8 &       11 \\
  1421$-$490 &     0.75 &      2462.5 &  456. &   72. &     114.3 &        5 \\
  1421$-$490 & $>$ 0.92 &  $<$  138.9 &  456. &   56. & $<$  18.1 &  $>$  11 \\
  1641$+$399 &     1.02 &        19.8 &  261. &  112. &      19.6 &       11 \\
  1642$+$690 &     0.95 &        68.6 &  352. &   79. &      22.8 &       10 \\
  1928$+$738 &     0.79 &      1102.5 &   80. &   34. &      84.8 &        6 \\
  2007$+$777 &     0.77 &      1823.9 &  102. &   16. &      48.7 &        8 \\
  2007$+$777 &     0.79 &      1115.7 &  102. &   22. &      51.5 &        8 \\
  2007$+$777 &     0.81 &       802.3 &  102. &   16. &      32.0 &        9 \\
  2007$+$777 &     0.81 &       810.3 &  102. &   16. &      31.0 &        9 \\
  2007$+$777 &     0.81 &       828.9 &  102. &   17. &      33.3 &        9 \\
  2123$-$463 &     0.85 &       423.0 &  540. &   57. &      18.6 &       11 \\
  2123$-$463 &     0.85 &       475.9 &  540. &   57. &      19.9 &       11 \\
  2255$-$282 &     0.92 &       124.5 &  430. &   58. &      18.9 &       11 \\
  2326$-$477 & $>$ 0.75 &  $<$ 2525.8 &  522. &   35. & $<$  41.9 &  $>$   8 \\
\enddata
\tablenotetext{a}{All quantities are defined as in Table~\ref{tab:beaming}.  Knots are defined in Table~\ref{tab:knotresults}.}
\end{deluxetable}

\scriptsize

\begin{table}
\begin{center}
\tablewidth{0pc}
\caption{Quasar Jet Orientations\tablenotemark{a} \label{tab:bends} }
\scriptsize
\begin{tabular}{rrrrcrrrrrrrr}
\tableline\tableline
Name & {PA$_{\rm kpc}$} & {PA$_{\rm pc}$} & {$\beta_{\rm app}$~~~~~} & Ref.\tablenotemark{b}
	& \multicolumn{3}{c}{{$\theta_{\rm kpc}$}} & \multicolumn{3}{c}{{$\zeta$\tablenotemark{c}}} 
	& {$\alpha_{rx}$\tablenotemark{d}} & {$\theta$\tablenotemark{d}} \\
{ } 	&	{ } 	&	{ } 	&	{ } 	&	{ } 	&	{min}	&	{mid}	&	{max}
	&	{min}	&	{mid}	&	{max}	& { } 	&  { } \\
\tableline
0106$+$013 & -175 &  -120 &  26.50 $\pm$ 3.90 &  1 &    3.6 &    4.2 &   11.7 &    1.8 &    2.5 &   11.2 &  \nodata & \nodata \\
0234$+$285 &  -20 &   -10 &  12.27 $\pm$ 0.84 &  1 &    1.8 &    4.9 &    7.9 &    1.0 &    1.2 &    5.2 &        0.86 &  9.1 \\
0707$+$476 &  -90 &   -20 &   6.74 $\pm$ 0.50 &  2 &   16.0 &   18.0 &   43.8 &    8.0 &   11.2 &   41.9 &  \nodata & \nodata \\
0745$+$241 &  -45 &   -60 &   7.90 $\pm$ 1.30 &  3 &    3.8 &    8.2 &   15.3 &    2.0 &    2.7 &   11.8 &   $>$  0.74 &  6.4 \\
0748$+$126 &  130 &   115 &  18.37 $\pm$ 0.82 &  1 &    1.8 &    3.5 &    6.7 &    1.0 &    1.2 &    5.1 &  \nodata & \nodata \\
0923$+$392 &   75 &   -78 &   4.29 $\pm$ 0.43 &  1 &   11.8 &   17.9 &   38.5 &    6.2 &    8.5 &   33.9 &   $>$  0.97 &  9.5 \\
0953$+$254 & -115 &  -120 &  12.40 $\pm$ 0.40 &  3 &    1.0 &    4.7 &    6.1 &    0.6 &    0.6 &    2.6 &  \nodata & \nodata \\
1055$+$018 &  180 &   -63 &  11.00 $\pm$ 1.00 &  1 &    9.4 &   10.6 &   28.7 &    4.8 &    6.6 &   27.4 &   $>$  0.92 & 10.1 \\
1055$+$201 &  -10 &    -5 &  10.00 $\pm$ 4.30 &  3 &    1.0 &    5.8 &    7.5 &    0.6 &    0.7 &    3.2 &        0.91 & 10.6 \\
1354$+$195 &  165 &   145 &   9.87 $\pm$ 0.06 &  2 &    4.0 &    7.0 &   14.9 &    2.0 &    2.8 &   12.5 &        0.95 & 10.1 \\
1502$+$106 &  160 &   120 &  14.80 $\pm$ 1.20 &  1 &    5.0 &    6.3 &   16.6 &    2.6 &    3.5 &   15.5 &  \nodata & \nodata \\
1641$+$399 &  -25 &   -95 &  19.27 $\pm$ 0.52 &  1 &    5.6 &    6.3 &   18.0 &    2.8 &    4.0 &   17.3 &        1.04 &  9.9 \\
1642$+$690 &  170 &  -166 &  16.00 $\pm$ 1.80 &  3 &    3.0 &    4.6 &   10.5 &    1.6 &    2.1 &    9.2 &        0.97 &  9.7 \\
1655$+$077 &  -50 &   -40 &  14.40 $\pm$ 1.10 &  1 &    1.4 &    4.2 &    6.8 &    0.8 &    1.0 &    4.4 &   $>$  0.97 &  9.9 \\
1823$+$568 &   90 &  -161 &  20.86 $\pm$ 0.49 &  1 &    5.2 &    5.9 &   16.8 &    2.6 &    3.7 &   16.2 &  \nodata & \nodata \\
1828$+$487 &  -40 &   -30 &  13.66 $\pm$ 0.39 &  1 &    1.6 &    4.4 &    7.1 &    0.8 &    1.0 &    4.6 &        0.91 &  6.5 \\
1928$+$738 & -170 &   170 &   8.43 $\pm$ 0.34 &  1 &    4.8 &    8.2 &   17.3 &    2.4 &    3.3 &   14.5 &        0.83 &  6.9 \\
2007$+$777 & -105 &   -95 &   0.82 $\pm$ 0.50 &  3 &    6.6 &   70.7 &   96.3 &    9.4 &   13.2 &   46.5 &        0.82 &  7.7 \\
2201$+$315 & -110 &  -140 &   7.88 $\pm$ 0.41 &  1 &    7.4 &   10.2 &   24.4 &    3.8 &    5.1 &   22.1 &  \nodata & \nodata \\
2230$+$114 &  135 &   160 &  15.41 $\pm$ 0.65 &  1 &    3.2 &    4.9 &   11.3 &    1.6 &    2.2 &    9.9 &  \nodata & \nodata \\
2251$+$158 &  -50 &   -60 &  14.19 $\pm$ 0.79 &  1 &    1.4 &    4.3 &    6.9 &    0.8 &    1.0 &    4.5 &        0.95 &  8.5 \\
2255$-$282 &  -70 &  -135 &   6.90 $\pm$ 1.00 &  2 &   15.2 &   17.1 &   42.1 &    7.6 &   10.6 &   40.2 &        0.95 & 10.5 \\
\tableline
\end{tabular}
\tablenotetext{a}{All angles are in degrees.  Position angles (PA) are defined relative to north, positive to the east.
  The min, mid, and max values give the minimum, 50\%, and 10\% probability points for the given angle.
  {\em Chandra} data for some sources were not yet available.}
\tablenotetext{b}{References for values of $\beta_{\rm app}$: (1) \citet{2009AJ....138.1874L};
(2) Lister et al.\ (in prep.);  (3) \citet{vlba}.}
\tablenotetext{c}{The quantity $\zeta$ is the angle between the pc-scale and kpc-scale jets in the frame of the quasar.
See eq.~\ref{eq:zeta}.}
\tablenotetext{d}{From Table~\ref{tab:beaming}.}
\end{center}
\end{table}

\end{document}